\documentclass[a4paper,fleqn,usenatbib]{mnras}

\usepackage[T1]{fontenc}
\usepackage{ae,aecompl}

\usepackage{float}

\usepackage{graphicx}	
\usepackage{amsmath}	
\usepackage{amssymb}	

\newcommand{\Hr}{\mathrm{H}}
\newcommand{\kms}{km~s$^{-1}$}
\newcommand{\Qd}{Q_\mathrm{D}^{\star}}
\newcommand{\dotM}{\dot{\mathrm{M}}}
\newcommand{\nel}{n$_{\mathrm{e}^{-}}$}
\newcommand{\nelo}{n$_{\mathrm{e}^{-}}^{0}$}

\title[HD~181327 debris ring]{Exocometary gas in the HD~181327 debris ring}

\author[S. Marino et al.]{ S. Marino$^{1,2,3}$\thanks{E-mail:
    s.marino@ast.cam.ac.uk}, L. Matr\`a$^{1}$, C. Stark$^{4}$,
  M. C. Wyatt$^{1}$, S. Casassus$^{2,3}$, G. Kennedy$^{1}$,
  \newauthor{D. Rodriguez$^{2,5}$, B. Zuckerman$^{6}$,
    S. Perez$^{2,3}$, W. R. F. Dent$^{7}$, M. Kuchner$^{8}$,}
  \newauthor{ A. M. Hughes$^{9}$, G. Schneider$^{10}$,
  A. Steele$^{11}$, A. Roberge$^{8}$, J. Donaldson$^{11}$ and
  E. Nesvold$^{12}$.}
\\
$^{1}$Institute of Astronomy, University of Cambridge, Madingley Road, Cambridge CB3 0HA, UK\\
$^{2}$Departamento de Astronom\'ia, Universidad de Chile, Casilla 36-D Santiago, Chile\\
$^{3}$Millennium Nucleus ``Protoplanetary Disks'', Chile\\
$^{4}$Space Telescope Science Institute, 3700 San Martin Dr, Baltimore, MD 21218, USA\\
$^{5}$Department of Astrophysics, American Museum of Natural History, Central Park West at 79th Street, New York, NY 10034, USA\\
$^{6}$Department of Physics and Astronomy, University of California, Los Angeles, CA 90095-1562, USA\\
$^{7}$Joint ALMA Observatory, Alonso de Córdova 3107, 763-0355 Vitacura, Santiago, Chile\\
$^{8}$Exoplanets and Stellar Astrophysics Laboratory, NASA Goddard Space Flight Center, Greenbelt, MD, USA\\
$^{9}$Van Vleck Observatory, Astronomy Department, Wesleyan University, 96 Foss Hill Drive, Middletown, CT 06459, USA\\
$^{10}$Department of Astronomy/Steward Observatory, The University of Arizona, 933 N. Cherry Ave., Tucson, AZ, 85721, USA\\
$^{11}$Department of Astronomy, University of Maryland, College Park, USA\\
$^{12}$Department of Terrestrial Magnetism, Carnegie Institution for Science, Washington, DC 20015, USA}

\date{Accepted 2016 May 16}

\pubyear{2015}

\begin{document}
\label{firstpage}
\pagerange{\pageref{firstpage}--\pageref{lastpage}}
\maketitle

\begin{abstract}

An increasing number of observations have shown that gaseous debris
discs are not an exception. However, until now we only knew of cases
around A stars. Here we present the first detection of $^{12}$CO (2-1)
disc emission around an F star, HD~181327, obtained with ALMA
observations at 1.3 mm. The continuum and CO emission are resolved
into an axisymmetric disc with ring-like morphology. Using a Markov
chain Monte Carlo method coupled with radiative transfer calculations
we study the dust and CO mass distribution. We find the dust is
distributed in a ring with a radius of $86.0\pm0.4$ AU and a radial
width of $23.2\pm1.0$ AU. At this frequency the ring radius is smaller
than in the optical, revealing grain size segregation expected due to
radiation pressure. We also report on the detection of low level
continuum emission beyond the main ring out to $\sim$200 AU. We model
the CO emission in the non-LTE regime and we find that the CO is
co-located with the dust, with a total CO gas mass ranging between
$1.2\times10^{-6}$~M$_{\oplus}$ and $2.9\times10^{-6}$~M$_{\oplus}$,
depending on the gas kinetic temperature and collisional partners
densities. The CO densities and location suggest a secondary origin,
i.e. released from icy planetesimals in the ring. We derive a
CO+CO$_{2}$ cometary composition that is consistent with
Solar system comets. Due to the low gas densities it is unlikely that
the gas is shaping the dust distribution.

\end{abstract}




\section{Introduction}


Recent surveys have shown that at least $\sim$20\% of nearby
solar-type stars (FGK) host Kuiper belt analogue debris discs
\citep{Hillenbrand2008, Bryden2009, Eiroa2013, Matthews2014}. From
multiwavelength observations we know that they are composed of dust
grains in a wide size distribution ranging from $\mu$m- to mm-sized
grains and collisional models require the presence of km-sized
planetesimals. These large bodies are a byproduct of planet formation
and continually replenish the dust population as the result of a
collisional cascade \citep[see][]{Wyatt2008}. The orbits of these
planetesimals can be drastically perturbed by the presence of planets,
and hence the disc density distribution can reveal a hidden planetary
system \citep[e.g.][]{Wyatt1999, Kuchner2003}. Thus, the study of
debris discs is an alternative method to characterize planetary
systems and the outcome of planet formation.  High resolution images
of debris discs have shown ring-like structures
\citep[e.g. HR~4796A,][]{Perrin2015}, warps \citep[e.g. $\beta$
  Pic,][]{Heap2000}, gaps or double ring structures
\citep[e.g. HD~107146,][]{Ricci2015} and eccentric rings
\citep[e.g. Fomalhaut,][]{Kalas2008}, suggesting the dynamical
presence of planets shaping the debris spatial distribution.


\begin{figure*}
  \includegraphics[trim=0.0cm 0.8cm 0.0cm 2.0cm, clip=true,width=1.0\textwidth]{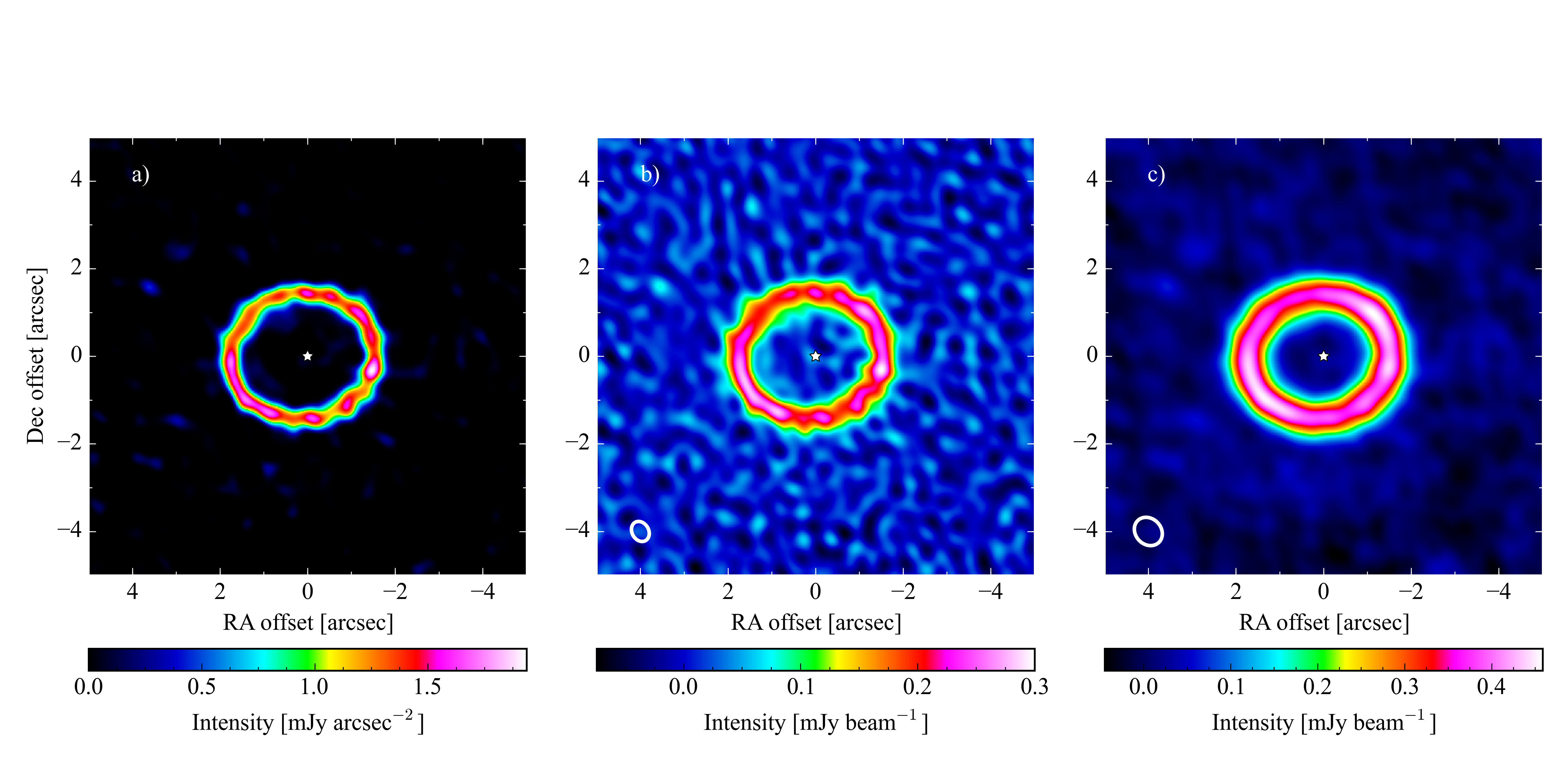}
  \caption{ALMA dust continuum maps at 220 GHz (Band 6). a) MEM
    non-parametric model (regularized with the maximum entropy
    method). b) Restored image adding residuals and convolving
    with a synthetic beam corresponding to \textit{Briggs} weighting
    ($0\farcs47\times0\farcs36$). c) Restored image adding
    residuals and convolving with a synthetic beam corresponding to
    \textit{natural} weighting ($0\farcs69\times0\farcs58$). The
    respective beams are represented by white ellipses. The x- and
    y-axes indicate the offset from the stellar position in R.A. and
    decl. in arcsec, i.e. north is up and east is left. The stellar
    position is marked with a white star.}
    \label{fig:ALMA_mem_restored}
\end{figure*}

Optical observations of discs can give insights about the distribution
of $\mu$m-sized dust grains at the bottom of the collisional cascade,
where stellar radiation and stellar winds become significant for the
dust dynamics \citep{Thebault2008}. On the other hand, dust thermal
emission at millimetre wavelengths is dominated by big grains
($\sim$0.1-10 mm) for which radiation forces are negligible
($\beta\equiv F_{\mathrm{rad}}/F_{g}\ll$1.0), tracing the location of
the parent bodies. Therefore, millimetre observations are fundamental
to study the distribution of the more massive planetesimals at the top
of the collisional cascade.

Moreover, in some of these systems significant amounts of gas has been
detected, especially in young debris discs that recently left the
protoplanetary disc phase. This gas can potentially impact the dust
dynamics and density distribution, producing structures that are
usually attributed to perturbing planets \citep{Lyra2013Natur}. The
origin of this gas is still under debate; however, recent observations
have allowed dynamical studies and thus a more detailed picture. This
has favoured a secondary origin scenario in the case of $\beta$
Pictoris \citep{Dent2014} and 49 Ceti \citep{Zuckerman1995,
  Zuckerman2012}, in which gas is released from icy bodies, e.g
through destructive collisions or photodesorption. In other systems
the gas seems to be primordial
\citep[e.g. HD~21997,][]{Kospal2013}. However, until now gas had only
been detected in seven debris discs, all of them around A
stars. \citep[see Table 5 in][for a complete list of debris discs with
  gas detections]{Moor2015}.

In this work we study the debris disc around the F5/6 main sequence
star HD~181327 \citep{Nordstrom2004, Torres2006}, member of the
$23\pm3$ Myr old $\beta$~Pic moving group \citep{Mamajek2014} and
located at a distance of $51.8\pm1.7$ pc \citep{vanLeeuwen2007}. The
star has an IR excess with a fractional luminosity of
$L_{\mathrm{IR}}/L_{\star}$ $\sim0.2$\% \citep{Lebreton2012} that
comes from dust thermal emission that has been marginally resolved at
different wavelengths: at 18.3 $\mu$m by \cite{Chen2008} and at 70
$\mu$m, 100 $\mu$m and 3.2 mm by \cite{Lebreton2012}. HST scattered
light observations resolved a debris ring of $\mu$m-sized grains with
a peak at $\sim90$ AU and 36 AU wide \citep{Schneider2006,
  Stark2014}. The latter study identified a possible radial dust size
segregation due to radiation pressure, based on variations of the
scattering phase function with radius. Moreover, they found
asymmetries that could be interpreted as an increase in the optical
depth in the west side of the ring, potentially due to either a recent
catastrophic disruption or warping of the disk by the ISM.

We present the first ALMA observations to study the dust continuum at
220 GHz and the CO gas distribution, constrain the location of
planetesimals, look for asymmetries that could give hints on the
origin of the asymmetric features observed by the HST, and study the
origin of the CO gas. In Sec. \ref{sec:obs} we present the
observations and imaging of the dust continuum and CO line
emission. In Sec. \ref{sec:model} we compare the observations with an
axisymmetric disc model for the dust continuum and CO (2-1), using a
Markov chain Monte Carlo (MCMC) method coupled with radiative transfer
simulations to sample the parameter space and estimate the debris
spatial distribution, CO gas mass and the disc orientation in the
sky. In Sec. \ref{sec:dis} we discuss the implications of the dust
continuum and CO observations, we compare them with the previous HST
observations, we derive a cometary composition based on the CO
observations, and we discuss scale height constrains from ALMA
observations of debris discs. Finally in Sec. \ref{sec:summary} we
summarise the main results and conclusions.

\section{ALMA observations and imaging}
\label{sec:obs}

HD~181327 was observed by ALMA in band 6 on 10 March 2014 as part of
the cycle 1 project 2012.1.00437.S. The total number of antennas was
26 with minimum and maximum projected baselines of 12 and 365 m,
respectively. The total time on source excluding overheads was 34
min. Three months later, on the 4 and 12 June, three new observation
runs using the same band were carried out corresponding to the cycle 2
project 2013.1.00523.S. The total number of antennas was 39 with a
minimum and maximum projected baselines of 15 and 650 m,
respectively. The total time on source excluding overheads was 129
min.

On the first run J1924-2914 was used as bandpass calibrator, while
Titan and J1819-6365 were used as primary flux and phase calibrators,
respectively. In the second project, J2056-4714 and J2056-472 were
used as bandpass calibrators, with the latter used as primary flux
calibrator too. We used J2009-4849 as phase calibrator. Calibrations
were applied using the pipeline provided by ALMA.

On both projects, the ALMA correlator provided 4 spectral windows
(spws), 3 exclusively dedicated to study the dust continuum with 128
channels and a total bandwidth of 2 GHz centered at: 212.9, 214.9 and
228.1 GHz in the first project, and 214.6, 216.6 and 232.5 GHz in the
second. The fourth spw was configured with a higher spectral
resolution of 244 KHz and total bandwidth of 937.5 MHz (3840
channels), centered at 230.5 GHz to target the $^{12}$CO (2-1)
transition at 230.538 GHz in both projects. We combine the two
datasets from both projects to achieve the most complete $u-v$
coverage and the highest signal-to-noise ratio (S/N) on the
reconstructed continuum image and CO line emission.


\subsection{Dust continuum}
\label{sec:obs_dust}
The image synthesis of the dust continuum was carried out using a
non-parametric least-squares modeling technique that incorporates a
regularization term called ``entropy'' from a family of Maximum
Entropy Methods (MEMs). Examples of usage of MEM for image synthesis
in Astronomy can be found in \citealt{Pantin1996, Casassus2006,
  Levanda2010, Casassus2013Natur, Warmuth2013, Coughlan2013,
  Marino2015mwc}. These deconvolved images "superresolve" the
interferometric data, as the entropy prior allows an extrapolation of
spatial frequencies beyond those sampled by the interferometer. We
call the whole algorithm \textit{uvmem} and the resulting images as
``MEM models''. The deconvolved image can then be ``restored'' by
convolving with a \textit{Clean} beam corresponding to
\textit{natural} or \textit{Briggs} weighting and adding the dirty map
of the residuals. The restored maps are comparable to standard Clean
images. While the resolution or point-spread-function on the
deconvolved image can vary at different locations, the resolution on
the restored images is well characterised by a synthetic beam of
$0\farcs47\times0\farcs36$ and $0\farcs69\times0\farcs58$,
corresponding to Briggs and natural weighting respectively.


In Figure \ref{fig:ALMA_mem_restored} we present the results of the
image synthesis on the continuum emission: (a) MEM model, (b) restored
image with Briggs weighting (Briggs map hereafter) and (c) restored
image with natural weights (natural map hereafter). To smooth the
artefacts due to thermal noise on the visibilities and characterize
the noise level in the image space, we perform a Monte Carlo
simulation bootstrapping the measured visibilities, i.e. adding a
random Gaussian noise according to the observed dispersion on each
base-line and repeating the image synthesis described above. The
presented images are the median of 200 iterations. We observe a noise
level of 0.022 mJy~beam$^{-1}$ on the Briggs map, while 0.015
mJy~beam$^{-1}$ on the natural map, estimated from the median absolute
deviation and from the rms on the dirty map of the residuals. The
total flux in the Natural map inside a ellipse of semi-major axis
$4\arcsec$ and oriented as the best-fit model (see
Sec. \ref{sec:model}) is $7.9\pm0.2$ mJy, the error which comes from
the image noise and does not take into account the error on the
absolute flux calibration ($\sim10\%$). This is consistent with the
trend of the spectral energy distribution (SED) between 170 $\mu$m
\citep[ISO]{Moon2006} and 3.2 mm \citep[ATCA]{Lebreton2012}, but too
low compared to the 0.05 Jy at 870 $\mu$m observed by LABOCA-APEX
\citep{Nilsson2009}, possibly due to a background galaxy in the
primary beam. Inside the primary beam we also detect two compact
sources with peak intensities of 0.14 and 0.08 mJy~beam$^{-1}$,
located at $8\arcsec$ with a PA=$9.5^{\circ}$ and $9\arcsec$ with a PA
of 210$^{\circ}$, respectively. These are probably background galaxies
(see Figure \ref{fig:bestmodel}) as the position of the southern
compact source is consistent with an edge-on galaxy observed in the
STIS/HST observations in 2011 \citep{Stark2014}.

Similar to previous observations of scattered light, the disc presents
a ring-like morphology. The bulk of the emission is radially confined
between $\sim50-125$ AU ($\sim1\farcs0-2\farcs4$) with a peak
intensity at $85.8\pm0.3$ AU obtained by fitting an ellipse to the
peak intensities along the ring of the MEM model. When we compare
these results with the HST observations we find that the ring is
slightly narrower and shifted inwards in the millimetre, while the
inner edge in scattered light ($82.3\pm1.1$) is close to the peak
radius in the millimetre. This is consistent with what would be
expected from grain size segregation due to radiation pressure,
previously suggested in this disc by \cite{Stark2014}. Radiation
pressure causes small dust grains to extend farther out in radius
compared with millimetre grains, shifting the density maximum of small
grains to larger radii (See discussion in Sec. \ref{dis:segregation}).

The disc emission appears consistent with an axisymmetric ring and
most of the observed intensity variations along the ring in the panels
of Figure \ref{fig:ALMA_mem_restored} and the small offset from the
star can be explained by the PA of the beam, the noise level and the
ALMA astrometric error (See Sec. \ref{dis:dust}). In
Sec. \ref{sec:model_dust} we compare with an axisymmetric model.



\begin{figure}
  \includegraphics[trim=0.4cm 2.0cm 1.5cm 3.0cm, clip=true, width=1.0\columnwidth]{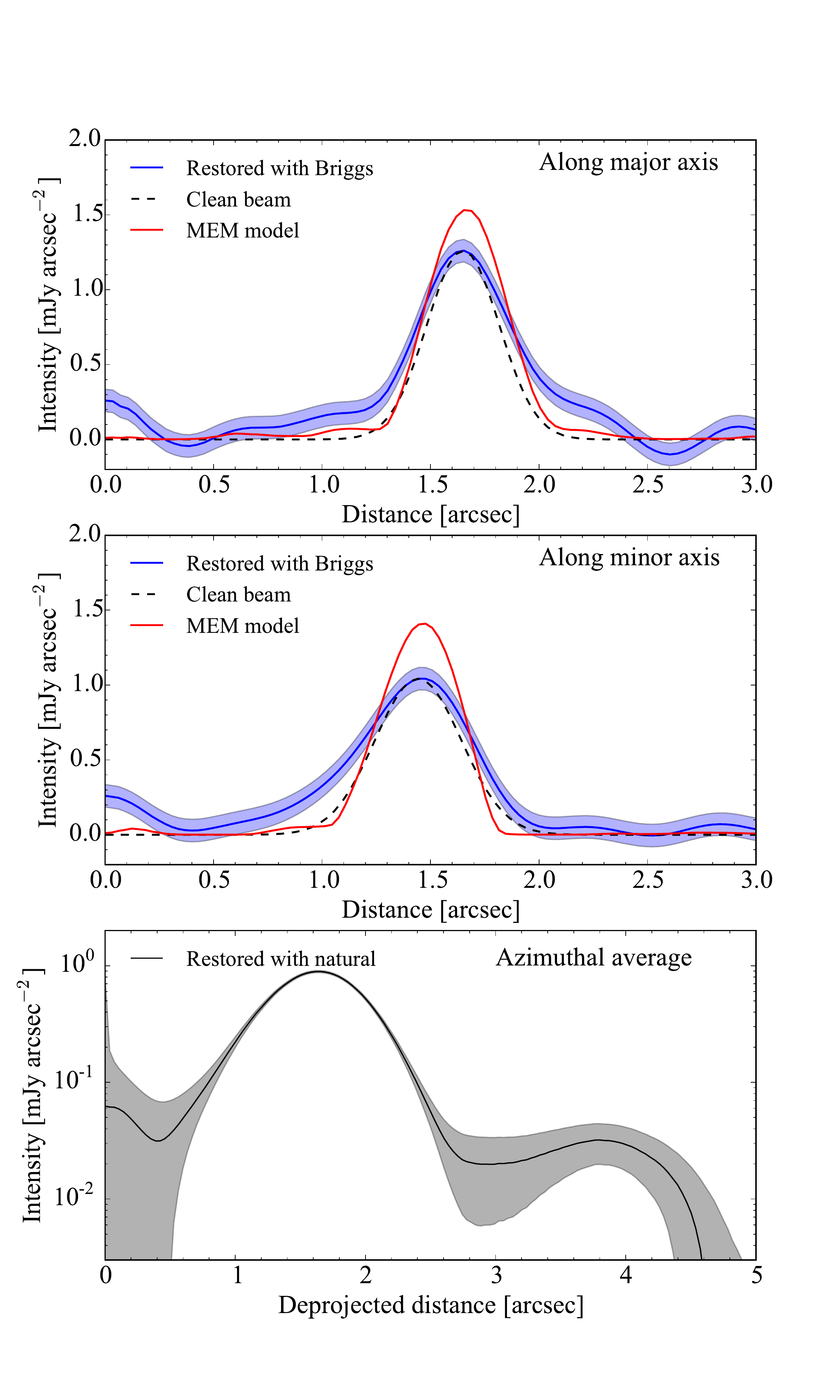}
  \caption{Intensity radial profiles of the dust continuum vs distance
    to the star along the major (top panel) and minor axis of the disc
    (middle panel). In the bottom panel we present the mean intensity
    at all azimuths vs the deprojected distance to the star. The red,
    blue, black dashed and black continuous lines represent the MEM
    model, Briggs map, projected Briggs Clean beam and natural map,
    respectively. The blue and grey areas represent the 68\% and
    99.7\% confidence regions, respectively. 
  }
    \label{fig:I_profile_Observed}
\end{figure}

In Figure \ref{fig:I_profile_Observed} we present intensity profiles
of the maps presented in Figure \ref{fig:ALMA_mem_restored}a and
\ref{fig:ALMA_mem_restored}b, along the ring's major and minor axis,
together with the deprojected mean intensity at all azimuths extracted
from Figure \ref{fig:ALMA_mem_restored}c. In the top and middle panels
the MEM model profile is represented by a red line, while the Briggs
map profile is represented in blue with blue shaded areas equivalent
to 1 $\sigma=0.022/\sqrt{2}$ mJy~beam$^{-1}$ (the factor $1/\sqrt{2}$
is due to the mean between the opposite sides of the disc). The
projected Briggs Clean beam is presented with dashed black lines. The
radial extent of the emission both along the major and minor axis is
wider than the beam, demonstrating that we can marginally resolve the
width of the ring. The full width half maximum (FWHM) of the ring is
$\sim0\farcs5$, which at a distance of 51.8 pc translates to $\sim25$
AU. A more accurate estimation is presented in
Sec. \ref{sec:model_dust}. The bottom panel shows that the disc
emission extends out to $\sim4\arcsec$ (200 AU) in either a second
ring or a halo component. The grey area in the bottom panel represents
the 99.7\% confidence region equivalent to 3
$\sigma=0.015/\sqrt{N_\mathrm{beams}}$ mJy~beam$^{-1}$ (the factor
$1/\sqrt{N_\mathrm{beams}}$ is due to the mean at all azimuth and the
number of independent points). In Figure \ref{fig:ALMA_mem_restored}c
this emission is also just visible as a broad ring around
$~4\arcsec$. To test the significance of this detection and whether
this could be a product of our image synthesis method, we produce
Clean images obtaining roughly the same results. Moreover, the second
component is recovered in the dirty map of the residuals when a
best-fit ring model is subtracted to the data (see
Sec. \ref{sec:model} for a description of the ring model). A brief
discussion about the origin of this emission and the dirty map of the
residuals is presented in Sec. \ref{sec:dis}.


\subsection{$^{12}$CO (2-1)}
\label{sec:co}

\begin{figure}
  \includegraphics[trim=0.5cm 0.5cm 0.4cm 0.0cm, clip=true,
    width=1.0\columnwidth]{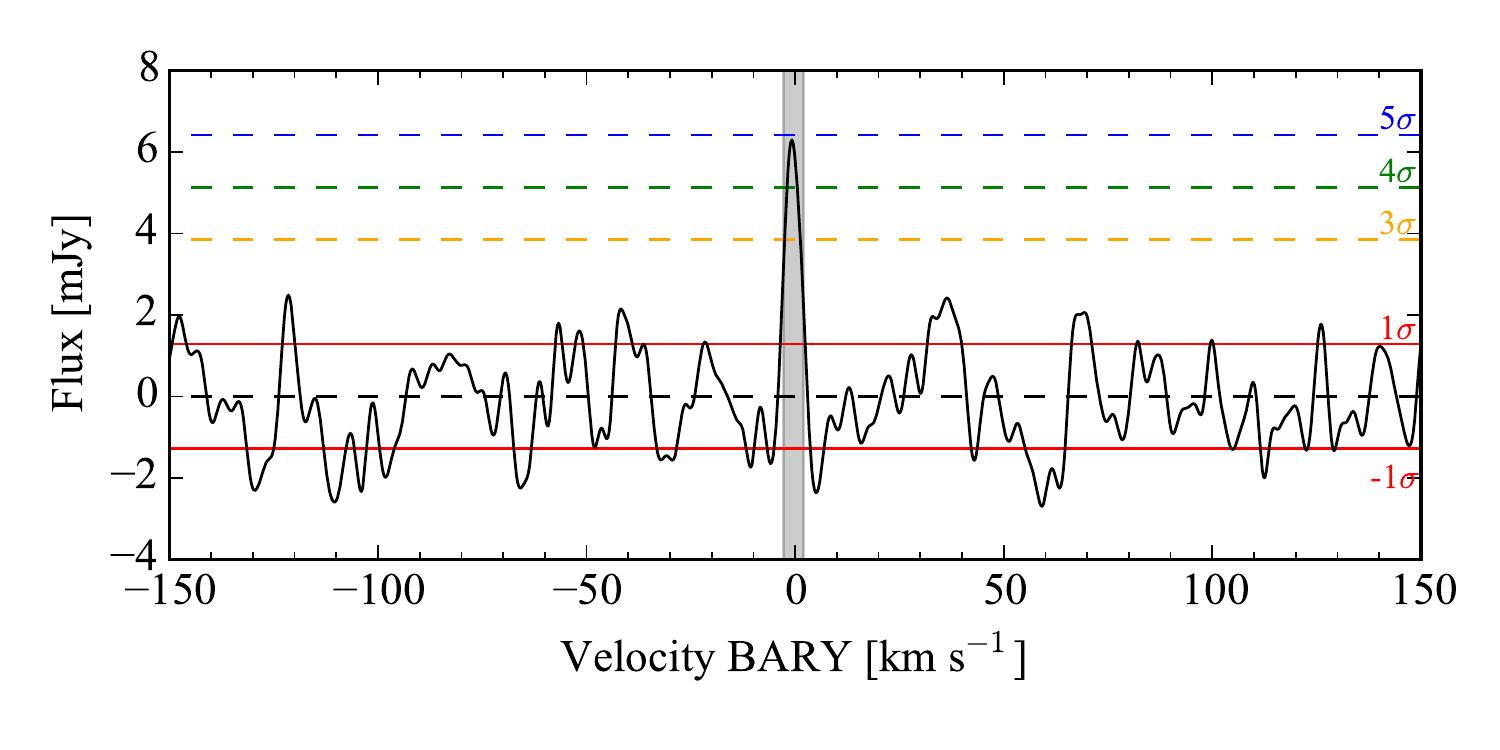}
  \caption{Continuum subtracted integrated spectrum inside an elliptic
    mask of minimum and maximum semi-major axis of $1\farcs4$ and
    $2\farcs1$, and oriented as the dust continuum ring. The original
    spectrum was smoothed with a Gaussian kernel with standard
    deviation of 4 channels. The horizontal lines represent $\pm1$, 3,
    4 and 5$\sigma$ levels. The grey region represents velocities
    between -2.8 \kms and 2.0 \kms where the S/N is maximised. The
    velocities represent the Doppler shift with respect to 230.538 GHz
    in the Barycentric reference frame.}
    \label{fig:spec}
\end{figure}


\begin{figure}
  \includegraphics[trim=0.5cm 0.5cm 0.4cm 0.0cm, clip=true,
    width=1.0\columnwidth]{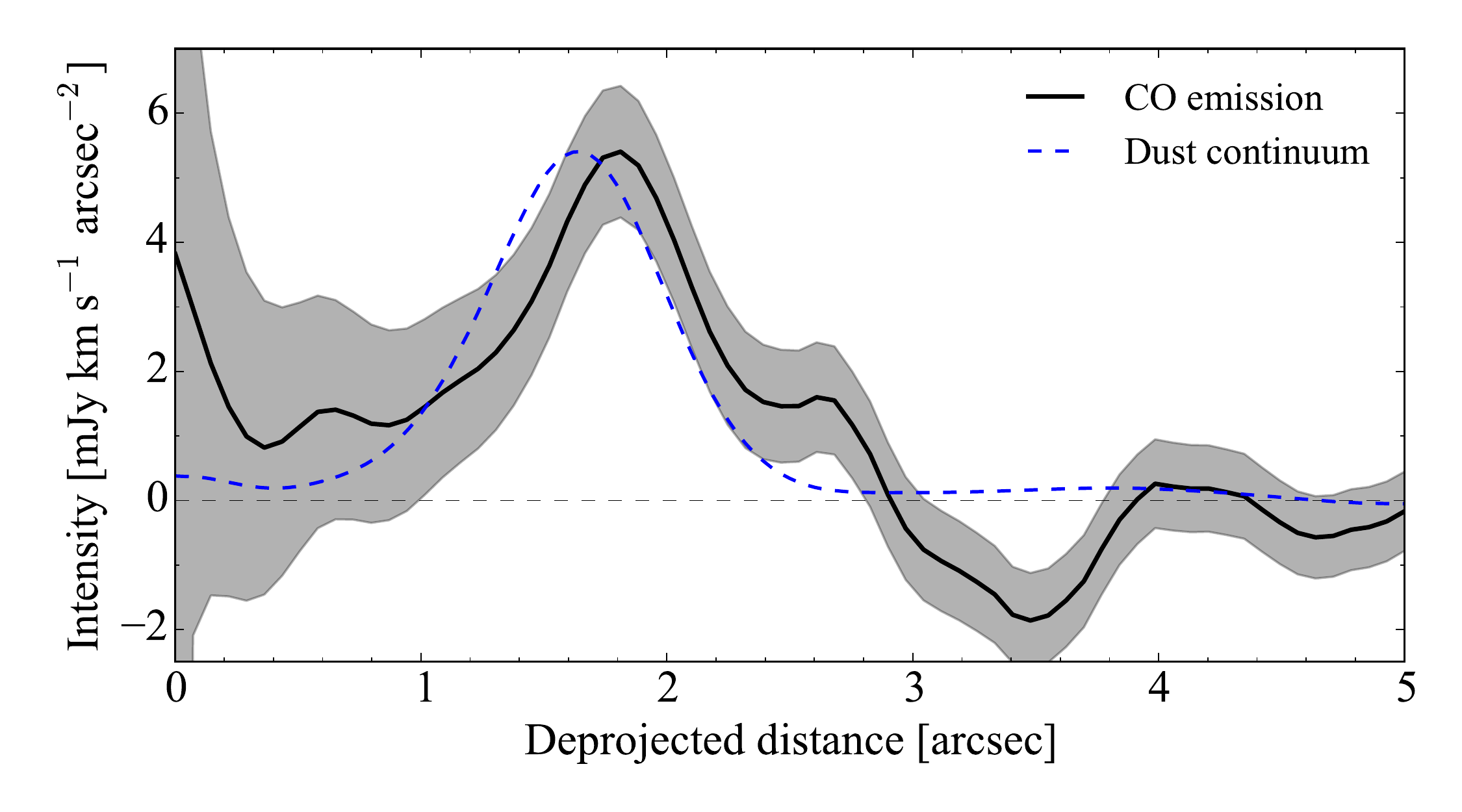}
  \caption{Mean intensity at all azimuths (black line) integrating
    from -2.8 \kms and 2.0 \kms with respect to 230.538 GHz, in the
    Barycentric reference frame. The profile is extracted from the
    dirty map with natural weights of the continuum subtracted
    visibilities. The grey area represents the 68\% confidence
    region. In blue we also present the mean intensity of the dust
    continuum presented in Fig. \ref{fig:I_profile_Observed} in an
    arbitrary scale.}
    \label{fig:Ir_co}
\end{figure}

To study the $^{12}$CO (2-1) transition line at 230.538 GHz (rest
frequency), we first subtract the continuum emission from the
visibilities fitting a polynomial of the first order at the
frequencies where no line emission is expected. The dirty map of the
continuum subtracted data does not show any significant emission above
$3\sigma$ which could be attributed to CO. With natural weights we
obtain a noise level of 1.0 mJy~beam$^{-1}$ per channel. To search for
any low level CO emission we compute the total flux inside an elliptic
mask and between a frequency or velocity range that we vary, both
spatially and in frequency, to maximise the S/N of the integrated
flux. The S/N is maximised inside an elliptic mask of minimum and
maximum semi-major axes of $1\farcs4$ and $2\farcs1$ (assuming the
same orientation and aspect ratio as the main ring) and a velocity
range of -2.8 to 2.0 \kms (Barycentric reference frame). These
parameters match the location of the main ring and the expected
Doppler shift due to Keplerian rotation, i.e. v$_{0}=0.1\pm0.4$
km~s$^{-1}$ \citep{Gontcharov2006} and v$_{\max}=1.9$ km~s$^{-1}$ for
M$_{\star}=1.36$~M$_{\sun}$ \citep{Lebreton2012}. The spectrum
extracted from the data cube inside the elliptic mask and smoothed
with a Gaussian kernel is presented in Figure \ref{fig:spec}. The
horizontal lines represent $\pm1$, 3, 4 and 5$\sigma$ levels. The line
peak is $\sim5\sigma$. The integrated flux where the S/N is maximised
is $30.1\pm 5.4$ mJy~km~s$^{-1}$ (5.6$\sigma$). Moreover, we find that
the CO emission in the south east and north west side of the disc are
consistent with being blue and red shifted, respectively. As the disc
north side is the brightest in scattered light due to forward
scattering, we can infer that it is also the closest side. This
implies that the disc is rotating clockwise in the sky.



In Figure \ref{fig:Ir_co} we present the integrated intensity over the
velocity range specified above and averaging at all azimuths. The grey
area represents the 99.7\% confidence region. The CO line peaks at
$90\pm4$ AU (error=$0.5\times$~beam semi-major axis $\times$ distance
/ (S/N)), close to the ring radius in the continuum, suggesting that
gas and dust are co-located. This favours a secondary origin scenario,
where the CO is being released from icy bodies in the disk. We study
the azimuthal profile of the emission along the ring and we find no
evidence for non-axisymmetry, similar to the dust continuum. In
Sec. \ref{sec:model_co} we study the CO gas distribution by modeling
the CO emission in the non-LTE regime, with different excitation
temperatures and electron densities which act as the main collisional
partner. Using an MCMC technique we find constraints on the
distribution of CO gas in the disc to quantify its similarity to the
dust distribution.


\section{Disc modelling}
\label{sec:model}
\subsection{Dust continuum}
\label{sec:model_dust}
To constrain the location of the millimetre-sized dust population, we
compare the observations with an axisymmetric debris disc model in the
Visibility space. The dust density distribution is parametrized as a
ring of radius $r_0$, with a Gaussian radial profile of width $\Delta
r$ (FWHM) and a Gaussian vertical profile with a scale height $\Hr$
(vertical standard deviation) or aspect ratio $h=\Hr/r$:
\begin{equation}
  \rho(r,z)=\rho_0 \exp\left[ -\frac{(r-r_0)^2}{2\sigma_r^2}
    -\frac{z^2}{2 \Hr^{2}} \right], \sigma_r=\frac{\Delta
    r}{2\sqrt{2\log2}}.
\end{equation}

For the dust optical properties we use a mass-weighted mean opacity of
astrosilicate grains with an internal density of 3.5 g~cm$^{-3}$
\citep{Draine2003}. We assume a Dohnanyi-like size distribution with a
power law index of -3.5, and minimum and maximum grain size
$a_{\min}=1.3 \mu$m and $a_{\max}=1.0$ cm. These values were roughly
estimated comparing the model and observed SED in an iterative
procedure. The mass-weighted absorption opacity
$\kappa_{\mathrm{abs}}=1.1$~cm$^{2}$~g$^{-1}$ at 1.3 mm, computed
using the ``Mie Theory'' code written by
\cite{BohreHuffman1983}. While $(a_{\min}, a_{\max})$,
$\kappa_{\mathrm{abs}}$ and the derived total dust mass are highly
dependent on our choice on the grain composition and size
distribution, these assumptions have very little effect on the derived
disc structure. A detailed study on the grain properties of this disc
can be found in \cite{Lebreton2012}. Synthetic images are computed
using
RADMC-3D\footnote{http://www.ita.uni-heidelberg.de/$\sim$dullemond/software/radmc-3d/}
\citep{RADMC3D0.39}, while the corresponding visibilities were derived
using our tool \textit{uvsim}. The free parameters in our model image
and visibilities are $r_0$, $h$, $\Delta r$, $M_{\mathrm{dust}}$, PA,
inclination ($i$) and RA- and decl-offset. The last two to account for
astrometric uncertainty and disc eccentricity.

We use a Bayesian approach to constrain the different parameters of
the ring model, sampling the parameter space to recover the posterior
distribution with the public python module \textit{emcee} that
implements the Goodman \& Weare's Affine Invariant MCMC Ensemble
sampler \citep{emcee}. The posterior distribution is defined as the
product of the likelihood function and the prior probability
distribution functions for each parameter, which we assume are
uniform. The likelihood function is defined proportional to
$\exp[-\chi^2/2]$, where $\chi^2$ is the sum over the squared
difference of the model and measured visibilities, divided by the
variance. In our priors, we impose a lower limit to $h$ equivalent to
0.03. This value is consistent with the minimum aspect ratio expected
in the absence of perturbing planets \citep{Thebault2009}.

\begin{figure}
  \includegraphics[trim=0.5cm 0.5cm 0.0cm 0.0cm, clip=true, width=1.0\columnwidth]{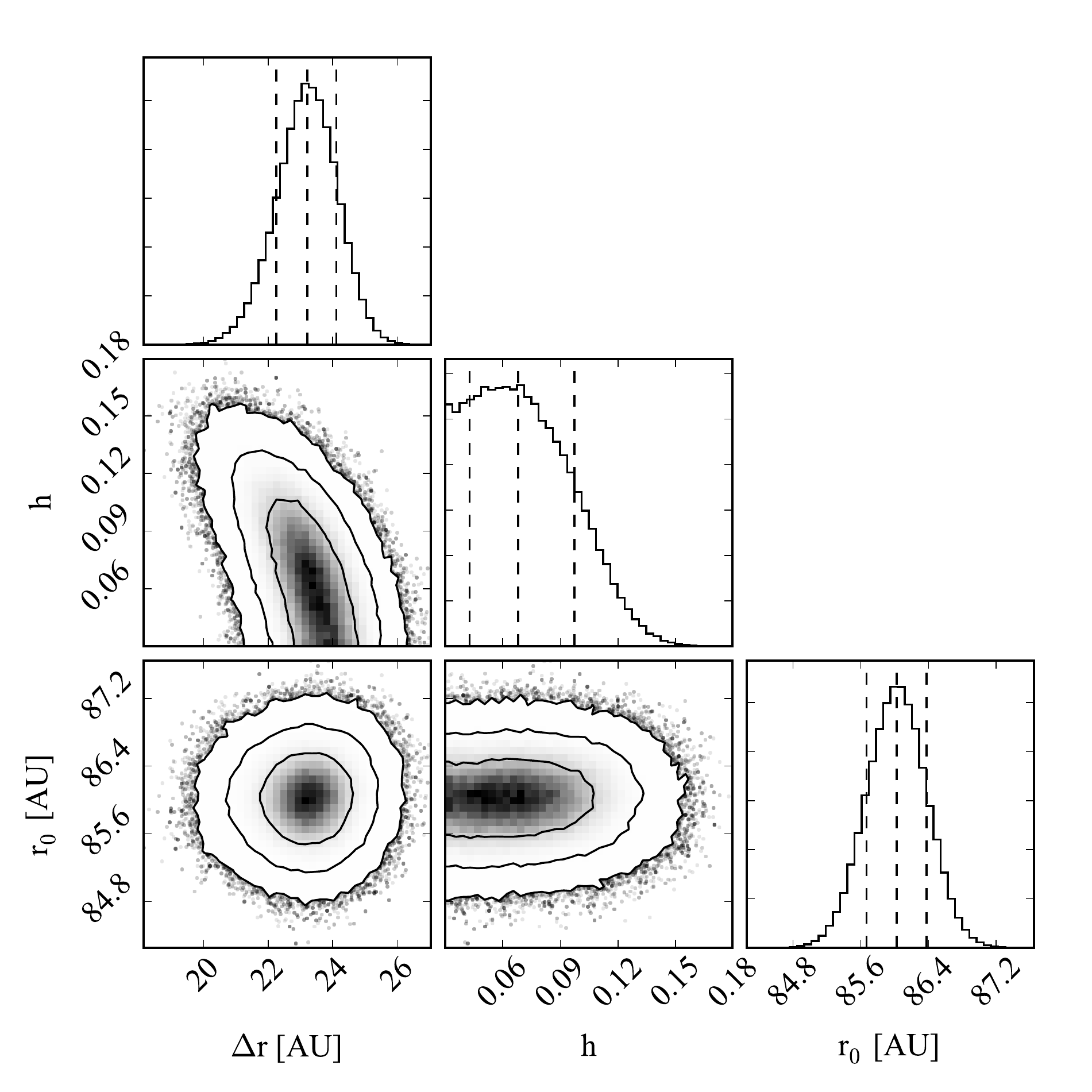}
  \caption{Posterior distribution of $h=\Hr/r$, $r_0$ and $\Delta
    r$. The marginalised distributions of $\Delta r$, $h=\Hr/r$ and
    $r_0$ are presented in the top, middle right and bottom right
    panel, respectively. The vertical dashed lines represent the
    16th, 50th and 84th percentiles. Left middle panel:
    marginalised distribution of $h$ and $\Delta r$. Left bottom
    panel: marginalised distribution of $r_0$ and $\Delta r$. Middle
    bottom panel: marginalised distribution of $r_0$ and $h$. Contours
    correspond to 68\%, 95\% and 99.7\% confidence regions and the
    black dots to points of the MCMC outside the 99.7\% confidence
    region. This plot was generated using the python module
    \textit{corner} \citep{cornerplot}. }
    \label{fig:mcmc}
\end{figure}

\begin{table}
  \centering
  \caption{Best fit values for the dust continuum model. Median
    $\pm$ uncertainty based on the 16th and 84th percentile of the
    marginalised distributions.}
  \label{tab:bestmodel}
  \begin{tabular}{cc} 
    \hline
    \hline
    Parameter & Best fit value \\
    \hline
    $M_{\mathrm{dust}}$ [M$_{\oplus}$] & $0.422\pm0.005$  \\
    $r_{0}$ [AU] & $86.0\pm0.4$ \\
    $\Delta r$ [AU] &  $23.2\pm1.0$ \\
    $h$ &  $0.07\pm0.03$  \\
    PA [$^{\circ}$] &  $98.8\pm1.4$ \\
    $i$ [$^{\circ}$] & $30.0\pm0.7$ \\
    RA offset [mas] & $96\pm5$  \\
    Dec offset [mas] &  $-42\pm5$ \\
    \hline
  \end{tabular}
\end{table}

The posterior distribution of $h$, $\Delta r$ and $r_0$ is presented
in Figure \ref{fig:mcmc}, while in Table \ref{tab:bestmodel} we
summarise the best fit values extracted from the marginalised
distribution. As we previously mentioned, the ring is marginally
resolved in the radial direction with an expectation value
$\langle\Delta r\rangle=23.2\pm1.0$ AU, where the uncertainty
represents the standard deviation derived from the marginalised
posterior distribution. The width of the ring can be well constrained
with the assumption of a Gaussian radial profile, but as it is of the
order of the beam size a detailed analysis of the radial structure of
the main ring is impossible with the current angular resolution. The
width of the ring is significantly larger than in the Fomalhaut ring
\citep{Boley2012}. This could be related to the age of the systems as
well as any possible planet- or star-disc interactions in the case of
Fomalhaut \citep{Faramaz2015, Shannon2014}. In agreement with the ring
radius derived in Sec. \ref{sec:obs}, we find $\langle
r_0\rangle=86.0\pm0.4$ AU.

On the other hand the scale height is not fully constrained. In
principle, it should be possible to derive $h$ from the ratio of the
ring's width or surface brightness along the major and minor axis of
the projected ring, causing $h$ and $\Delta r$ to be correlated. In
addition to this, in our case the estimation of $h$ is limited as our
observations only marginally resolve the ring's width.  However, we
can put a 99.7\% confidence upper limit of 0.14 for $h$. Both upper
limit and best fit value (see Table \ref{tab:bestmodel}) are
consistent with the upper limit of 0.11 from \cite{Stark2014} and 0.09
from \cite{Schneider2006}. The same figure shows that $h$ and $\Delta
r$ are anticorrelated (see Sec. \ref{dis:h}). In Sec. \ref{dis:h} we
discuss about determining $h$ in future observations and in other
discs. The derived total dust mass in the main ring is $0.442\pm0.005$
M$_{\oplus}$. This value is highly dependant on our choice of dust
internal density (3.5 g~cm$^{-3}$) and for our assumed dust
size distribution scales as $(a_{\max}/1~\mathrm{cm})^{0.5}$, noting
that $a_{\max}$ is unconstrained. Moreover, the uncertainty on the
dust mass quoted above does not include the uncertainty on the
absolute flux calibration, or the uncertainty on the stellar
distance. Including these systematics, the uncertainty should be
$\sim12\%$.


Our best axisymmetric model has
$\chi^2_{\mathrm{red}}\equiv\chi^2/(N-\nu-1)=1.54$, where $N$ and
$\nu$ are the number of independent measurements (visibilities) and
free parameters and equal to 74888 and 8, respectively. When we
compare the $\chi^2$ of our best axisymmetric model with the MEM model
we find a difference of 0.2\%. Hence, we conclude the continuum
emission is consistent with an axisymmetric ring.


\subsection{$^{12}$CO (2-1)}
\label{sec:model_co}

To derive the CO gas distribution we follow a procedure similar to the
dust continuum modeling in the previous Section. We fit the data with
an axisymmetric disc of gas with a density distribution parametrized
similar to the dust density distribution, but with a fixed vertical
aspect ratio $h$ of 0.07 and fixed disc orientation, that corresponds
to the best fit of the dust continuum (see Table
\ref{tab:bestmodel}). The gas is in Keplerian circular orbits with a
fixed systemic radial velocity of 0.1 km~s$^{-1}$ in the Barycentric
frame. Instead of simulating visibilities as in the dust continuum
analysis, we fit the data with our gas models in the image space. The
model images produced with RADMC-3D at different frequencies (tracing
different radial velocities) are convolved with the dirty beam and
then compared directly with the dirty map of the CO data, both
corresponding to natural weights. This method is analogous to
comparing model visibilities with gridded visibility data. The
likelihood function is defined proportional to $\exp[-\chi^2/2]$,
where $\chi^2$ is the sum over the image and frequency space of the
squared difference between the model and observed dirty map, divided
by the variance, and taking into account the number of independent
beams in the image and the correlation between adjacent channels in
ALMA data. The free parameters in the MCMC are the CO gas mass
(M$_\mathrm{CO}$), ring radius ($r_0$) and FWHM of the ring ($\Delta
r$).

However, the CO emission is not only constrained by the density
distribution. \cite{Matra2015} showed that local thermodynamic
equilibrium (LTE) does not necessarily apply in the low gas density
environments of debris discs. Depending on the gas kinetic temperature
and collisional partner densities, the derived gas mass can vary by
orders of magnitude. Thus, to model the $^{12}$CO (2-1) emission it is
necessary to include non-LTE effects, i.e to solve the population
levels including (de-)excitations through radiative process, where the
CMB photons and dust thermal emission dominate the radiation field at
230 GHz, as well as through collisions. The similarity between the
dust and gas distribution showed in Sec. \ref{sec:co} suggest that the
CO is of secondary origin, e.g. produced by collisions between icy
planetesimals that release volatile species such as CO or H$_2$O. This
implies that the main collisional partner of CO molecules are
electrons produced by carbon ionization and H$_{2}$O, which we neglect
as is a much less efficient collider compared to electrons. However,
we stress that the derived CO gas distribution and total mass (in both
the extremes of radiation dominated and LTE) are independent of the
specific collisional partner, because it is the density of the
partners that matters. That is, a different partner would yield a
different range of values on the x-axis, but the same CO masses. We
computed the CO-e$^{-}$ collisional rates using expressions described
by \cite{Dickinson1975}. In Sec. \ref{sec:coorigin} we discuss the
origin of the CO gas. We also use as input our best fit model of the
dust continuum to compute the dust contribution to the radiation field
at 230 GHz.

We consider different kinetic temperatures ranging from half to twice
the dust temperature derived in the previous section (50 K at the ring
radius). The electron density is defined as
\nel(r)=\nelo($r$/86~AU)$^{-1.1}$, where the power law index was
assumed equal to the one derived for $\beta$ Pic
\citep{Matra2016inprep}. We vary \nelo between
10$^{-2}$-$10^{7}$~cm$^{-3}$ to cover from the radiation-dominated
regime to LTE. Scattered light images suggest that the north east side
of the disc is the closest as it is brighter in scattered light
because of forwards scattering; thus, if the CO is in Keplerian
rotation we should be able to determine the direction at which the CO
moves in the sky. Hence, we consider models of the disc rotating
clockwise as well as anticlockwise. We find a best fit with the gas
rotating clockwise on the sky. The best model is five times more
likely than the best anticlockwise model and 300 times more likely
than no CO emission. In Sec. \ref{dis:dust} we discuss the
implications for the observed asymmetries in scattered light. Below,
we concentrate only on the clockwise case.

\begin{figure}
  \includegraphics[trim=0.0cm 0.0cm 1.0cm 1.0cm, clip=true,
    width=1.0\columnwidth]{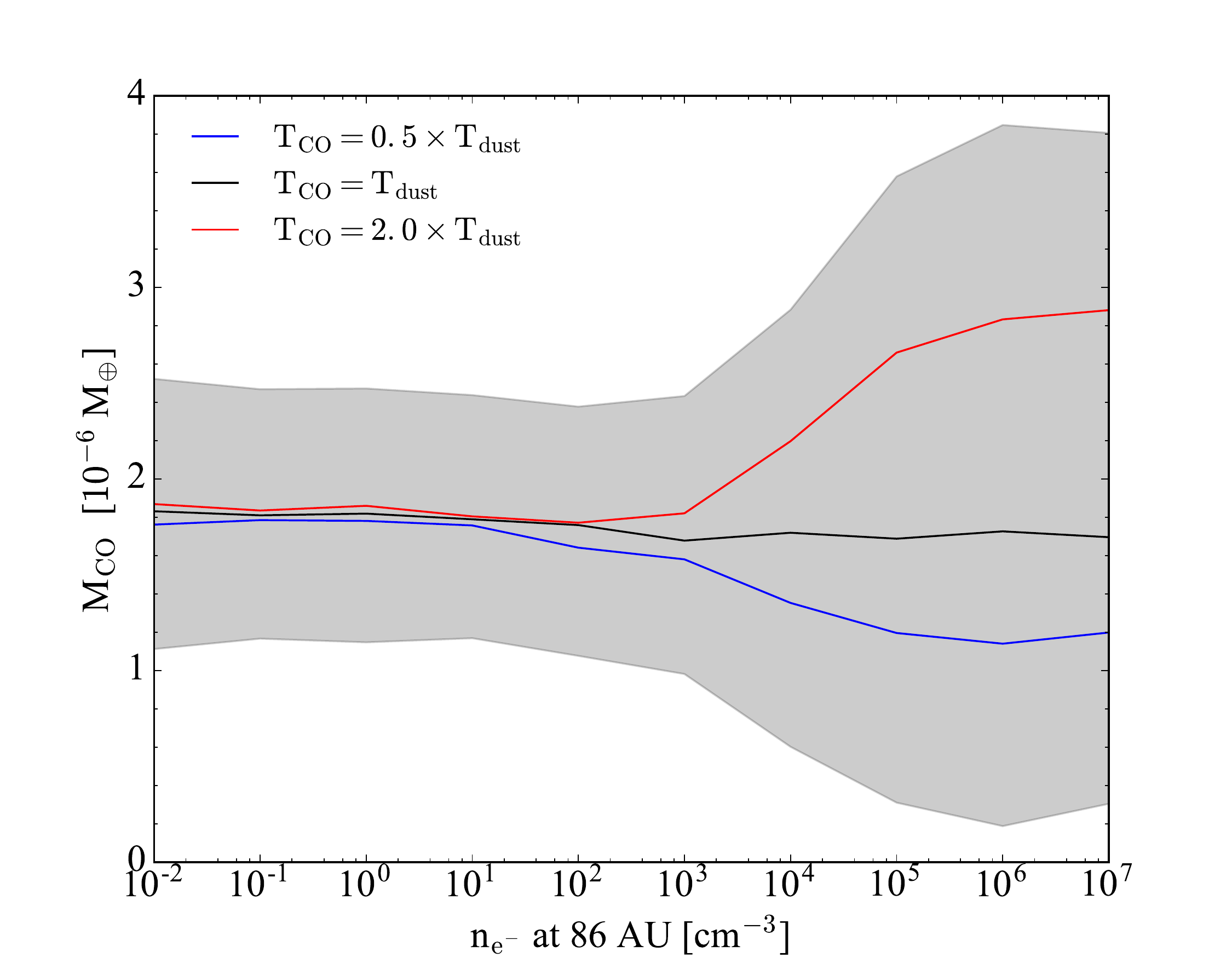}
  \caption{Best fit CO gas mass values obtained from our MCMC analysis
    using different kinetic temperatures and electron densities
    (\nelo, main collisional partner). The blue, black and red lines
    correspond to kinetic temperatures equal to 0.5, 1 and 2.0 times
    the dust temperature in the disc, respectively. The grey area
    represents the 68\% confidence region with gas temperatures
    ranging between $0.5-2.0$ times the dust temperature.}
  \label{fig:m_co}
\end{figure}

\begin{figure}
  \includegraphics[trim=0.0cm 0.0cm 0.5cm 0.0cm, clip=true,
    width=1.0\columnwidth]{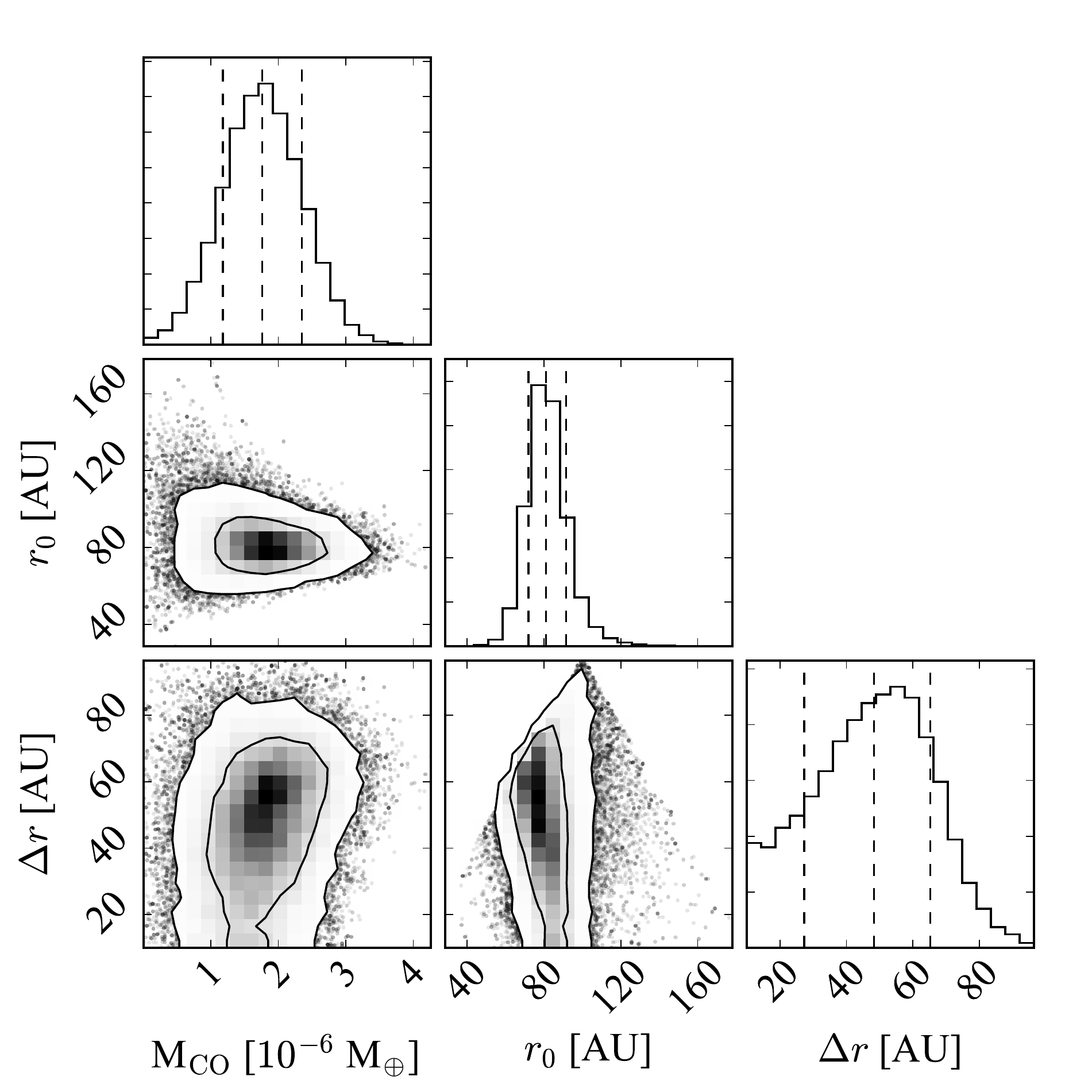}
  \caption{Posterior distribution of M$_\mathrm{CO}$, $r_0$ and
    $\Delta r$ of the CO disc. The marginalised distributions of
    M$_\mathrm{CO}$, $r_0$ and $\Delta r$ are presented in the top,
    middle right and bottom right panel, respectively. The vertical
    dashed lines represent the 16th, 50th and 84th percentiles. Left
    middle panel: marginalised distribution of M$_\mathrm{CO}$ and
    $r_0$. Left bottom panel: marginalised distribution of $\Delta r$
    and M$_\mathrm{CO}$. Middle bottom panel: marginalised
    distribution of $r_0$ and $\Delta r$. Contours correspond to 68\%,
    and 95\% confidence regions and the black dots to points of the
    MCMC outside the 99.7\% confidence region. This plot was generated
    using the python module \textit{corner} \citep{cornerplot}. }
    \label{fig:mcmc_co}
\end{figure}

In Figure \ref{fig:m_co} we present the CO gas masses derived with our
MCMC-modeling technique for different kinetic temperatures and
electron densities. We find that our best fit value of M$_\mathrm{CO}$
can vary due to the uncertainty in the excitation temperature between
$1.2\pm0.4\times10^{-6}$ - $2.9\pm 0.9\times10^{-6}$~ M$_{\oplus}$ in
the worst case (LTE), while it is very well constrained around
$1.8\pm0.6\times10^{-6}$~M$_{\oplus}$ in the radiation dominated
regime (low \nel). Recent ALMA observations of the $\beta$~Pic disc
suggest \nelo$\sim10^{2}$~cm$^{-3}$ \citep{Matra2016inprep}, which for
HD~181327 corresponds to a scenario very close to the radiation
dominated regime. In Figure \ref{fig:mcmc_co} we present the posterior
distributions, assuming \nelo$=10^{2}$~cm$^{-3}$ and equal gas and
dust temperatures. We find M$_\mathrm{CO}=1.8\pm0.6
\times10^{-6}$~M$_{\oplus}$, $r_0=81^{+10}_{-9}$~AU and $\Delta
r=48^{+17}_{-21}$~AU. This model has a total flux of $\sim$ 18
mJy~km~s$^{-1}$, a CO peak density of $\sim$0.15 cm$^{-3}$ and peak
optical depth of $\sim0.007$. We also notice that when both gas and
dust temperature are equal, the difference in the derived CO mass
between the radiation dominated regime and LTE is negligible because
the radiation in the ring is actually dominated by the dust thermal
emission rather than the CMB, as it was found to be the case in the
Fomalhaut debris ring \citep{Matra2015}. We also find that the derived
values of $r_0$ and $\Delta r$ are independent of \nelo and the gas
kinetic temperature. Table \ref{tab:mcmc_co} summarises the best fit
values of the CO distribution.

The radius of the CO ring matches with the dust ring, confirming what
we found averaging the data in Sec. \ref{sec:co}, that gas and dust
are co-located. Given the derived dust and CO densities it is unlikely
that the CO is self shielded enough to avoid the photodissociation due
to the ISM radiation field, which occurs in a timescale of $\sim120$
yr \citep{Visser2009} (see discussion in
Sec. \ref{sec:coorigin}). This implies that the CO must be replenished
and produced in the main ring; therefore, we conclude that the CO is
of secondary origin and released by icy bodies. In
Sec. \ref{sec:coorigin} we discuss its origin. We also find that the
CO ring could have a radial width similar to the dust distribution,
although it is not very well constrained.

\begin{table}
  \centering
  \caption{Best fit values for the CO model. Median $\pm$ uncertainty
    based on the 16th and 84th percentile of the marginalised
    distributions.}
  \label{tab:mcmc_co}
  \begin{tabular}{cc} 
    \hline
    \hline
    Parameter & Best fit value \\
    \hline
    M$_\mathrm{CO}$ [M$_\oplus$] & $1.2\pm0.4\times10^{-6}$ - $2.9\pm0.9\times10^{-6}$ \\
    $r_{0}$ [AU] & $81^{+10}_{-9}$ \\
    $\Delta r$ [AU] &  $48^{+17}_{-21}$ \\
    \hline
  \end{tabular}
\end{table}

\begin{figure}
  \includegraphics[trim=0.5cm 0.5cm 0.4cm 0.0cm, clip=true,
    width=1.0\columnwidth]{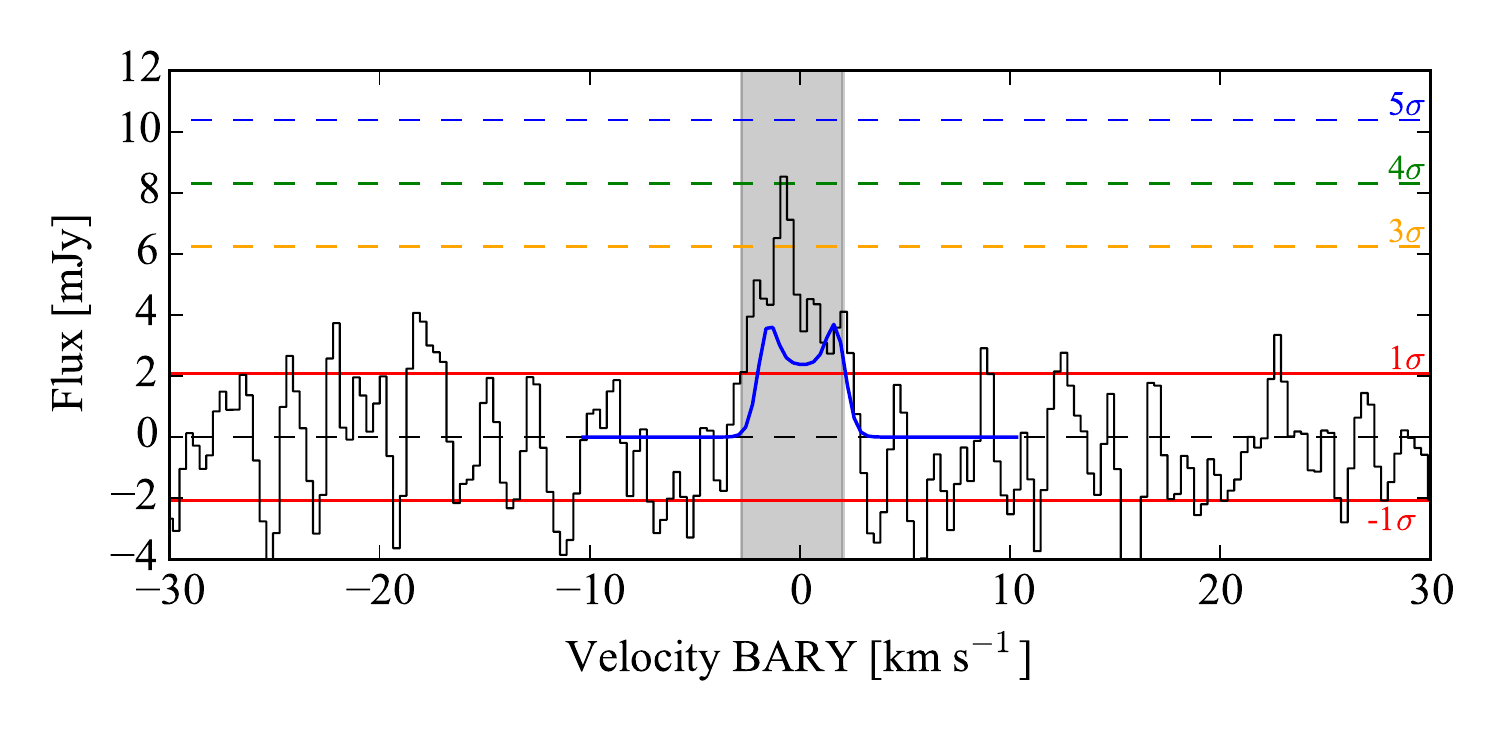}
  \caption{Blue line: model spectrum. Black line: continuum subtracted
    integrated spectrum inside an elliptic mask of minimum and maximum
    semi-major axis of $1\farcs4$ and $2\farcs1$, and oriented as the
    dust continuum ring. The horizontal lines represent $\pm1$, 3, 4
    and 5$\sigma$ levels. The grey region represents velocities
    between -2.8 \kms and 2.0 \kms. The velocities represent the
    Doppler shift with respect to 230.538 GHz in the Barycentric
    reference frame.}
    \label{fig:specmodel}
\end{figure}

In figure \ref{fig:specmodel} we compare the model spectrum with the
spectrum extracted from the dirty map only integrating over the south
east half of the disk for negative radial velocities, and over the
north west half of the disk for positive radial velocities. With this
approach we obtain an integrated flux of $21.8\pm4.3$ mJy~km~s$^{-1}$,
matching our best fit model and still consistent with the integrated
flux presented in Sec. \ref{sec:co} ($30.1\pm5.4$ mJy~km~s$^{-1}$)
within the uncertainties. The model spectrum matches roughly the
observed CO line profile.

\begin{figure*}
  \begin{center}
  \includegraphics[trim=0.5cm 0.5cm 0.0cm 0.0cm, clip=true, width=1.0\textwidth]{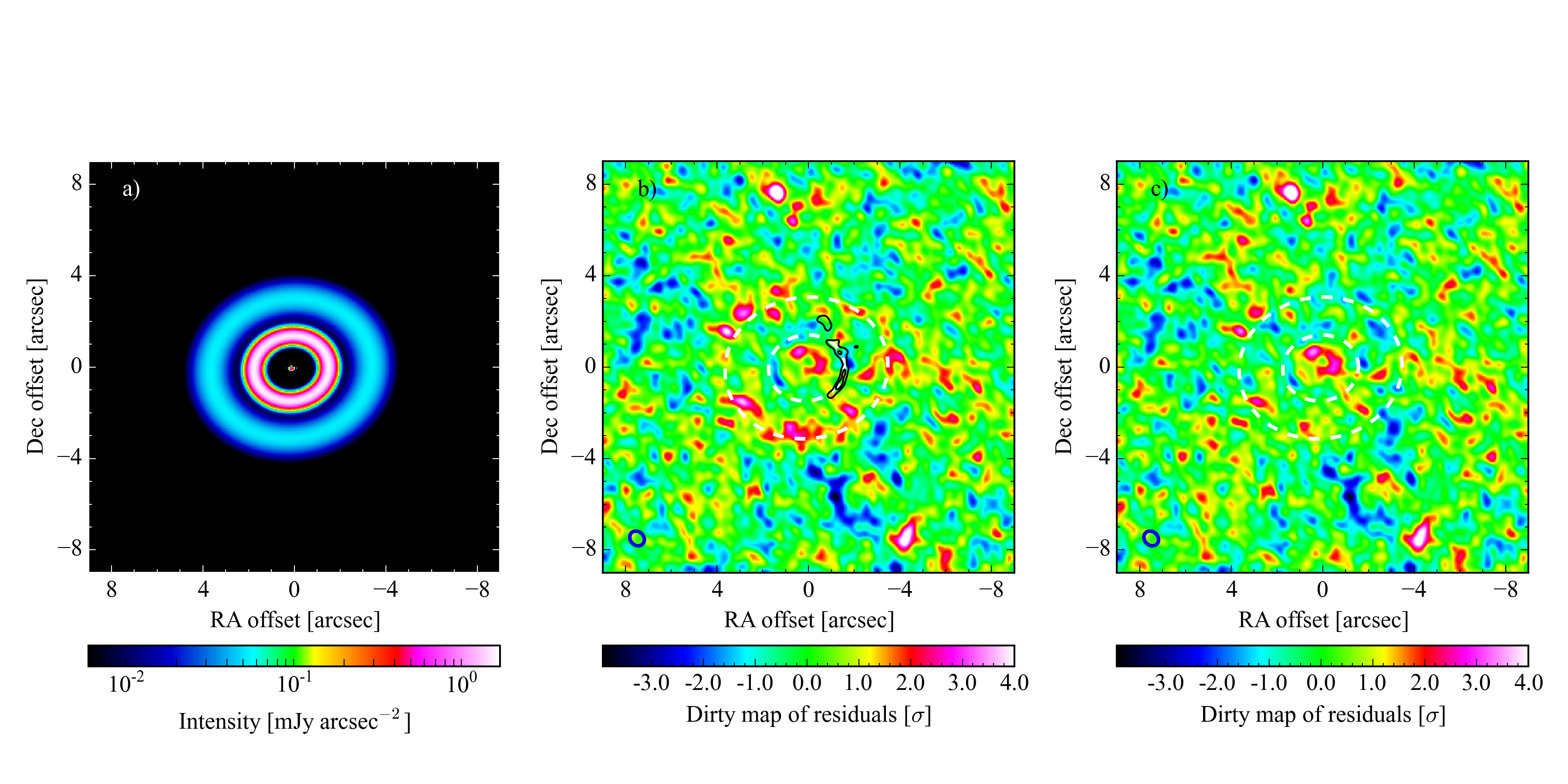}
  \caption{ a) Synthetic model image of the best fit model with two
    rings at 220 GHz. b) Dirty map of the residuals of the best fit
    model with a single ring corresponding to natural weighting. The
    black contours represent the optical depth deviations from a
    uniform disc derived from \textit{HST} images \citep[][left panel
      Figure 1]{Stark2014}. c) Dirty map of the residuals of the best
    fit model with two rings corresponding to natural weighting. The
    white dashed contours in panels b) and c) correspond to radius of
    the main and second ring of the best fit two rings model. The
    Clean beam is represented by a blue ellipse
    ($0\farcs67\times0\farcs58$) in the bottom left corner. The x- and
    y-axes indicate the offset from the stellar position in R.A. and
    decl. in arcsec, i.e. north is up and east is left.}
    \label{fig:bestmodel}
\end{center}\end{figure*}

\section{Discussion}
\label{sec:dis}
\subsection{Axisymmetry}
\label{dis:dust}

In Sec. \ref{sec:model} we analysed the observations assuming an
axisymmetric disc, however \cite{Stark2014} found large scale
asymmetries consistent with either a recent catastrophic disruption of
a large body or disc warping due to interactions with the ISM. Figure
\ref{fig:bestmodel} shows the best axisymmetric ring model image (a),
the dirty map of the residuals when the best model is subtracted from
the observed visibilities using natural weighting (b), and when the
best double ring model is subtracted (c, see Section below). The map
of residuals of a single ring is consistent with pure thermal noise
without any peak intensity greater than 3$\sigma$ along the main ring
or where scattered light observations suggested an increase in the
optical depth (black contours). The same image shows the two compact
sources at $\sim8\arcsec$ N and $9\arcsec$ SW from the star described
in Sec. \ref{sec:obs_dust}. We can put an upper limit on any
fractional enhancement of emission in the millimetre of $\sim$10\%,
which is lower than the asymmetries derived in scattered light
(10-42\%). This can be translated to a mass upper limit of any dust
density enhancement of $5\times 10^{-3}$ M$_{\oplus}$~beam$^{-1}$
($\sim3$ Pluto mass), assuming the dust composition and grain size
distribution ($a_{\max}=1.0$ cm) described in
Sec. \ref{sec:model}. This mass upper limit scales roughly as
($a_{\max}$/1 cm)$^{1/2}$ with the Dohnanyi-like size distribution, as
noted in Sec. \ref{sec:model_dust}.

In a giant collision scenario, the small dust produced from the
collision would initially form a trailing outward-propagating spiral
structure due to radiation pressure that would orbit the star
\citep{Kral2013}. After one orbit the fragments would collide again in
the ``pinch point'' and would continually produce new debris in this
region where the density is higher. The small dust produced by
collisions would form a leading outward-propagating spiral structure
from the pinch point as it is affected by radiation pressure
\citep{Jackson2014}, and we would expect the millimetre-sized dust
distribution to be narrower at the collision point, increasing the
surface brightness. In addition, such a collision would produce a CO
excess that would extend from the collision point, along the ring in
the direction of motion as far as it can spread in $\sim120$ yr at
which point is photodissociated, as the disc is optically thin,
similar to the case of $\beta$ Pictoris \citep{Dent2014, Jackson2014}.

Such features are not present in the millimetre continuum, nor in the
CO emission. In Sec. \ref{sec:co} and \ref{sec:model_co} we found that
the CO and, thus, all the debris rotates clockwise in the sky. This
has strong implications for the interpretation of the asymmetries in
scattered light that extend outwards in an anticlockwise direction,
forming a trailing spiral. The giant collision scenario could only
explain the asymmetry if the collision has occurred in the last
$4\times10^3$ yr or 5 orbits; thus, very unlikely. An alternative
explanation of the asymmetry in scattered light could be that the disc
is being warped by ISM interactions as suggested by \cite{Stark2014}
or that the excess is being produced by a large body releasing small
dust \citep[e.g ][]{Rappaport2014}, in which case the asymmetric
structure would form a trailing spiral of small grains that would
orbit the star in clockwise direction.


\subsection{Extended emission}
Even though we model the disc as a single ring, we detect emission
that extends at least to $4\arcsec$ (see
Sec. \ref{sec:obs_dust}). This emission is recovered both at the east
and west side of the disc with no significant differences, and using
our image synthesis method as well as using the Clean
algorithm. Moreover, it appears in the residuals presented in Figure
\ref{fig:bestmodel}b. Assuming the same dust composition and grain
size distribution as in the original ring, we model it with a second
component in the dust density distribution. We find that this emission
can be fitted by a power law surface density distribution, that starts
in the main ring with a surface density 25 times lower and that
decreases as $r^{-1}$. An alternative is to add a second ring, in
which case we find that the observations are best fitted with a second
ring of radius $\sim185$ AU, FWHM$\sim76$ AU and peak surface density
25 times lower than the maximum of the main ring. In both alternative
models we obtain a total flux of $8.6\pm0.4$~mJy and a dust mass of
$\sim0.57\pm0.04$ M$_{\oplus}$, assuming the same grain size
distribution for the two components, an assumption which, as noted
above, has significant influence on the derived dust mass. The flux
uncertainty does not consider the uncertainty on the absolute flux
calibration ($\sim10\%$). Figure \ref{fig:bestmodel}c shows the
residuals when the double ring model is subtracted, obtaining a map
with pure thermal noise.  The origin of this emission could be dust on
eccentric orbits produced in the main ring from eccentric
planetesimals, or even primordial dust from the protoplanetary disc
phase as this is a young system ($\sim23$ Myr). This extended
component could be related to the change in slope of the derived
surface density near 150 AU in scattered light \citep[see figure 9
  in][]{Stark2014}. Deeper ALMA observations at a different wavelength
or even ACA observations coupled with detailed modeling are necessary
to conclude about its origin. If the dust surface density distribution
has a local minimum between $2\arcsec$ and $4\arcsec$, this could be
evidence of a perturbing planet orbiting in the gap or on a highly
eccentric orbit \citep{Pearce2015doublering}.

\subsection{Dust size segregation}
\label{dis:segregation}
From our MCMC analysis and the fit of an ellipse to the intensity
maxima along the ring, we find that its radius is $86.0\pm 0.4$,
significantly smaller than at optical wavelengths
($r_0=90.5\pm1.1$). On the other hand, the distribution of small and
millimetre grains overlaps. The main ring of millimetre grains extends
roughly from 63 to 110 AU, while the small dust is found roughly from
70 AU and beyond 200 AU. The difference in peak radius of $4.5 \pm
1.2$ AU reveals grain size segregation. Due to radiation pressure,
small dust grains released from larger bodies on circular orbits
should be put on eccentric orbits with larger semi-major axes that
depend on the grain sizes. The net effect is that the larger grains
traced at millimetre wavelengths should remain in almost circular
orbits at the radius of the parent planetesimal belt, while the
spatial distribution of small grains shifts to larger radii
\citep[see][for detailed modelling]{Thebault2008}. The same
segregation has been proposed for other debris discs, e.g. in AU
Microscopii \cite{Strubbe2006} suggested a parent planetesimal belt at
a specific radius to explain the surface brightness profile from
scattered light images. This was later corroborated by millimetre
observations \citep{MacGregor2013}.

\subsection{Collisional timescales}
\label{dis:coll}
Rough estimates of the time scale at which mass is being lost
($\dotM$) from planetesimals in the steady state collisional cascade
can be made under the following assumptions: (1) the particles in the
disc have small mean eccentricities and inclinations ($e$ and $I$)
equal to 0.05; (2) the relative velocities between particles is equal
to v$_\mathrm{K}(1.25e^2+I^2)^{1/2}$ \citep[valid for Rayleigh
  distributions of $e$ and $I$, ][]{Lissauer1993, Wetherill1993}; (3)
planetesimal strengths ($\Qd$) independent of size and equal to 230
J~kg$^{-1}$ \citep[appropiate for km-sized weak ice bodies,
][]{Benz1999, Wyatt2002}. Constant $\Qd$ leads to a universal particle
size distribution with a power law index of -3.5, consistent with our
assumptions in Sec. \ref{sec:model_dust}. Under these assumptions and
using equations 15 and 16 from \cite{Wyatt2008} we find $\dotM \sim 3$
M$_{\oplus}$~Myr$^{-1}$. By equating the collisional lifetime of
bodies of different size with the age of the system (23 Myr), we also
infer a maximum planetesimal size in the collisional cascade
($\mathrm{D}_c$) of at least 1.8 km in diameter. Larger
planetesimals may be present, but they would not yet have collided,
thus they do not contribute to the collisional cascade. The
collisional lifetime scales with the size of the body, $\mathrm{D}_c$,
roughly as $(\mathrm{D}_c/1.8$ km$)^{0.5}$. We can then extrapolate
this to obtain a collisional lifetime of mm-sized grains of
$24\times10^{3}$ yr ($\sim34$ orbits).

Note that this collision timescale is broadly consistent with equation
16 in \cite{Wyatt1999} given the different assumptions for these
calculations. This emphasises that the collisional lifetime of the
mm-sized grains is set by the cross-sectional area in small grains,
which is relatively well constrained from the observations, and is not
dependent on the details of the collisional lifetimes of the largest
objects discussed above.


\subsection{Eccentric ring?}
In our analysis we also found that the ring center is offset by
$96\pm5$ mas in RA and $-42\pm5$ mas in decl with respect to the phase
center, which was centered at the expected stellar position at
epoch. However, the offset is within the astrometric rms of ALMA
($\sim0\farcs1$; private communication with ALMA helpdesk). An upper
limit for the eccentricity of the ring can be estimated using a 3
$\sigma$ astrometric error of $0.3\arcsec$. This results in an
eccentricity upper limit of 0.18, consistent with the eccentricity of
$0.02\pm 0.01$ measured by \cite{Stark2014}.


\subsection{CO origin}
\label{sec:coorigin}
We found that the CO is co-located with the dust, with no evidence for
non-axisymmetry and with a radial extension consistent with the width
of the main ring, although not well constrained. Given the CO derived
mass and limits, we can estimate a vertical and radial column density
of about $10^{13}$ and $10^{14}$ cm$^{-2}$ depending on the gas
kinetic temperature and electron densities (see
Sec. \ref{sec:model_co}), which implies a CO self shielding
coefficient of $\sim0.6-0.8$ \citep{Visser2009} (low self-shielding)
and a photodissociation timescale of $\sim150-200$~yr due to the
interstellar radiation field \citep[see details in
][]{Matra2015}. Even if we assume that the CO is primordial and there
is 10$^4$ times more H$_{2}$ than CO, the column density is still too
low to shield the CO. Hence, the CO gas must have been produced
recently and is probably continually replenished through destructive
collisions or photodesorption of icy planetesimals, i.e. it has a
secondary origin. Furthermore, given that the CO cannot exist for an
orbital period at these distances (680 yr at 86 AU), the majority of
detected CO must be produced roughly axisymmetrically throughout the
disk ring.

If we compare the derived CO mass with other debris discs with CO gas
detected and of secondary origin \citep[49~Ceti and $\beta$~Pic,][,
  respectively]{Dent2005, Dent2014}, we find that for HD~181327
M$_\mathrm{CO}$ is at least an order of magnitude lower. Moreover, the
CO/dust mass ratio is between $3-7\times10^{-6}$ in HD~181327, two
orders of magnitude lower compared to $\beta$~Pic (CO/dust mass ratio
$\sim 3\times10^{-4}$). The difference could be in the host star as
49~Ceti and $\beta$~Pic are A stars, which could naturally favour the
release of volatiles due to stronger radiation environments.

\subsection{Cometary composition}

In Sec. \ref{dis:coll} we derived the mass loss rate from
planetesimals ($\dotM$) and in Sec. \ref{sec:coorigin} we determined
that the CO must be of secondary origin. Assuming that the mass of CO
present in gas phase is in steady state, we can derive the CO ice mass
fraction $f_\mathrm{CO}$ of planetesimals as a function of $\dotM$ and
the photodissociation timescale of CO $\tau_\mathrm{co}$. In steady
state we expect $f_\mathrm{CO}\times
\dotM=\mathrm{M}_\mathrm{CO}\times\tau_\mathrm{co}^{-1}$, thus
\begin{equation}
f_\mathrm{CO}= 3.4\times 10^{-3} \left(\frac{\mathrm{M}_\mathrm{CO}}{1.8\times10^{-6}~\mathrm{M}_{\oplus}}\right)\left(\frac{\dotM}{3~\mathrm{M}_{\oplus}~\mathrm{Myr}^{-1}}\right)^{-1}\left(\frac{\tau_\mathrm{co}}{175~\mathrm{yr}}\right)^{-1}.
\end{equation}
This value could vary from 3$\times 10^{-3}$ to 6$\times 10^{-3}$ due
to systematic uncertainties in M$_\mathrm{CO}$ and
$\tau_\mathrm{co}(\mathrm{M}_\mathrm{CO})$. Moreover, the value of
$\dotM$ is highly dependant on $\Qd$ which could vary between 200 and
10$^4$ J~kg$^{-1}$ for km-sized bodies, making f$_\mathrm{CO}$ range
between 3$\times 10^{-3}$ to 0.11. Another big uncertainty is that an
important fraction of CO could be also a product of CO$_2$, released
from icy bodies and that photodissociates in a shorter timescale of
$\sim30$ yr \citep{Hudson1971, Lewis1983}, or produced by CO$_{2}$ ice
photodesorption. Thus, the CO ice fraction above can be interpreted as
CO+CO$_2$ ice mass fraction of planetesimals. The net effect is that
$f_\mathrm{CO+CO_2}$ is probably between 0.3\%-16\% which is
consistent with the CO+CO$_2$ abundances in Solar system comets
\citep[CO+CO$_{2}$ mass fraction of 3-27\%,][]{Mumma2011}. If the
planetesimals have an ice to rock fraction similar to the Solar System
of about unity, we can extrapolate and obtain an (CO+CO$_2$)/H$_2$O
ice abundance ratio between 0.4-18\%. We stress that the CO and CO$_2$
production rate does not depend on the unknown mechanism that is
releasing CO, as it will always be limited by the destruction rate of
icy solids that sublimate CO or expose an icy surface.

\subsection{Dust-gas interactions}
The origin of the ring-like morphology of the HD 181327 disk is
unclear. \cite{Lyra2013Natur} showed that dust-gas interactions can be
non-negligible in debris discs and produce instabilities that shape
the dust and gas distribution in narrow ring-like structures. These
instabilities arise if: the dust stopping time $\tau_{f}$ is in the
range of 0.1-10 $\times 1/\Omega_\mathrm{K}$, with $\Omega_\mathrm{k}$
the Keplerian rotation frequency, and dust to gas ratio
$\varepsilon\lesssim1$. To ascertain if this could explain the dust
distribution in the HD~181327 debris disc we need to estimate $\tau_f$
and $\varepsilon$. A reasonable value for the stopping time as a
function of the grain size can be obtained using the CO gas mass
derived in our modeling,
i.e. M$_\mathrm{CO}=1.2\times10^{-6}-2.9\times10^{-6}~\mathrm{M}_{\oplus}$,
and if we assume: (1) the gas is dominated by CO, H$_2$O and their
photodissociaton products; (2) the CO/H$_2$O ice mass ratio in
planetesimals derived above (0.4-18\%); (3) the C/CO abundance ratio
is $\sim100$, the same as in $\beta$~Pic \citep{Roberge2006}. Under
these assumptions we find a total gas mass that could vary between
$6\times10^{-4}$ and 0.04 $\mathrm{M}_{\oplus}$.


The stopping time defined as
$\tau_s=m\mathrm{v}_\mathrm{rel}/F_\mathrm{drag}$ can be estimated
considering the Epstein drag force that gas exerts on dust grains. We
approximate the dust grain velocities with the same expression for the
relative velocities presented in Sec. \ref{dis:coll}. The gas drag
force depends on the gas density, gas kinetic temperature ($T_k$) and
the grain size ($a_d$). The first can be derived assuming a gas
density distribution proportional to the dust density distribution,
while the second is assumed to be close to the mm-dust dust grain
temperature in the ring ($\sim$50 K). Finally we obtain
\begin{equation}
  \tau_{\mathrm{s}} \simeq 3\times10^3 \ \left(\frac{\mathrm{M}_\mathrm{gas}}{0.04 \ \mathrm{M}_{\oplus} }\right)^{-1} \ \left(\frac{a_d}{\mathrm{1 \ mm}}\right) \left(\frac{T_k}{50 \ \mathrm{K}}\right)^{-1/2} \Omega_K^{-1}.
\end{equation}

Thus, for mm-sized particles $\tau_{\mathrm{s}}\gtrsim3\times10^{3}
\Omega^{-1}$, much longer than the necessary to trigger the
instability. Moreover, the stopping time is even longer than the
collisional lifetime calculated in Sec. \ref{dis:coll} ($\sim34$
orbits). Only for grains at the bottom of the collisional cascade,
with sizes $\lesssim10$ $\mu$m, the stopping time is in the range
where the instability could be triggered. The dust-to-gas mass ratio
of grains smaller than 10 $\mu$m is of the order of unity. However,
given that this instability does not apply to the mm-sized grains, and
the distribution of micron-sized grains is consistent with being
derived from the mm-sized grains, with any differences in their
distribution attributable to radiation pressure without the need for
the photoelectric instability, we conclude that this instability does
not necessarily play a significant role in this disk. Furthermore, the
collisional lifetime of $\mu$m-sized grains can be obtained using the
fractional luminosity of the disk \citep{Wyatt1999}, which translates
to $\sim$5 orbits, while the timescale for the instability to occur is
tens of orbital periods \citep{Lyra2013Natur}.


\subsection{Scale height}
\label{dis:h}

We showed that ALMA observations can be useful to measure or constrain
the vertical mass distribution in discs, even for non highly inclined
discs. Moreover, the dust thermal emission is not dependant on the
scattering phase function, which hinders scale height estimates from
scattered light observations. In the case of HD~181327, we found an
upper limit for $h$ consistent with previous observations. As
$h$ is directly related to the mean orbital inclination relative to
the disk midplane, measurements of it at (sub)millimetre wavelengths
can be a powerful tool for identifying perturbing massive bodies,
e.g. planets on inclined orbits stirring the disc vertically.

\begin{figure}
  \begin{center}
  \includegraphics[trim=0.0cm 0.5cm 1.5cm 2.0cm, clip=true,
    width=1.0\columnwidth]{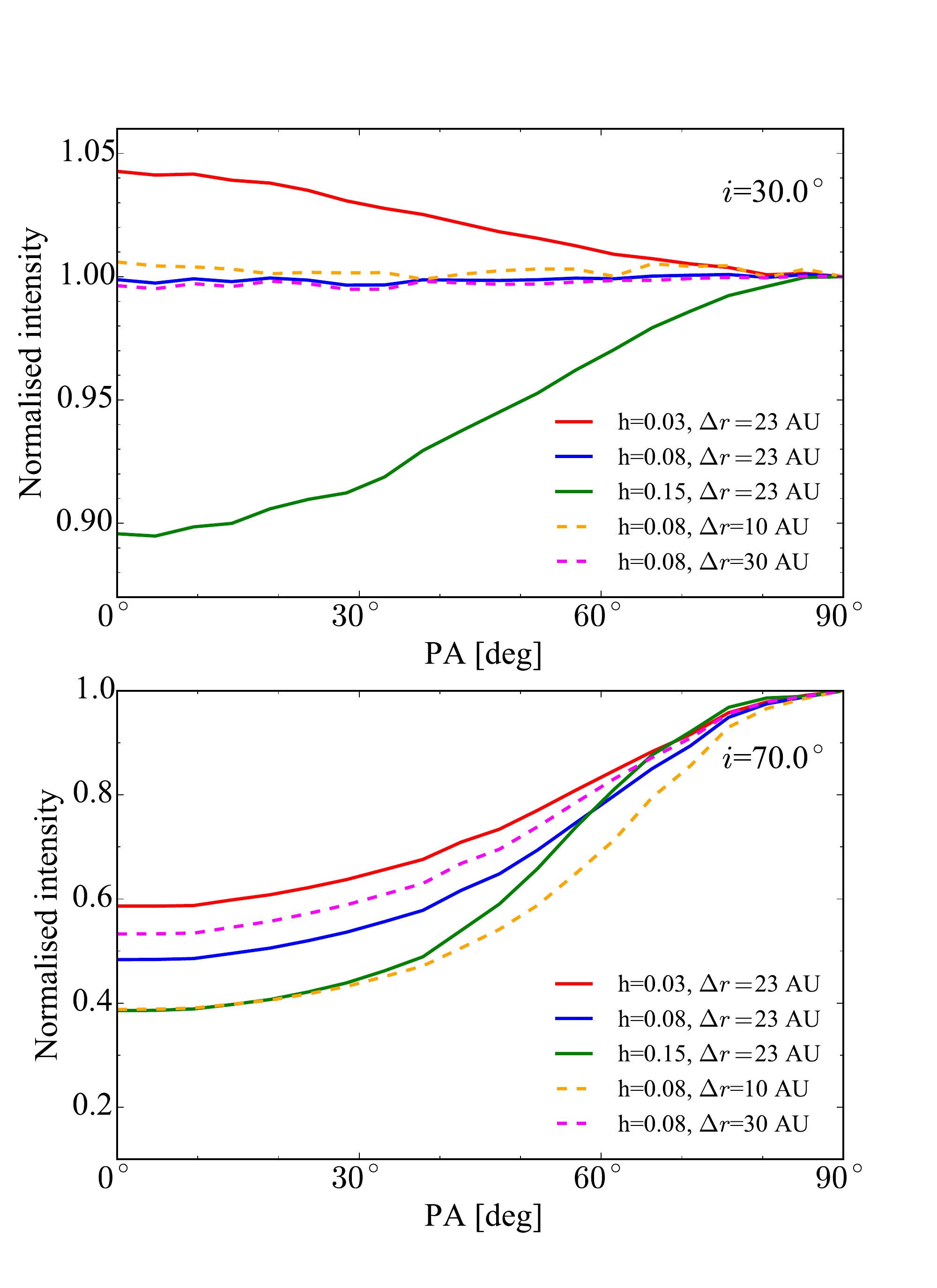}
  \caption{Intensity profiles along the ring projected ellipse on sky
    obtained from simulated Clean images without noise and using our
    best fit ring model. The values are normalised to the intensity at
    the ring semi-major axis (PA=$90^{\circ}$). The upper panel
    corresponds to a disc inclined by $30^{\circ}$, while the lower to
    $70^{\circ}$. In red, blue and green lines we present three
    different profiles for disc aspect ratios of 0.03, 0.08 and 0.15
    with $\Delta r$ fixed to 23 AU, while in dashed lines we present
    profiles varying $h$ and $\Delta r$.}
    \label{fig:Iphi}
\end{center}\end{figure}

To illustrate why ALMA observations can constrain $h$ in narrow discs,
we study the brightness profile at a given radius of model disc
observations with different $h$, $\Delta r$ and $i$. For a given
surface density profile, different $h$'s result in different azimuthal
intensity profiles due to changes of the optical depth in the line of
sight. For a narrow ring, higher values of $h$ make the disc brighter
at the ansae compared to the regions close to the disc
minor-axis. This is shown in Figure \ref{fig:Iphi} where we plot
intensity profiles, normalized at the disc ansa (PA=90$^{\circ}$),
versus position angle obtained from simulated ALMA cycle 3
observations. We used as input image our best-fit ring model and we
vary $h$ and $\Delta r$. The equivalent angular resolution of the
simulated observations is $\sim0\farcs25$. The upper panel corresponds
to a disc inclination similar to HD~181327 ($30^{\circ}$), while the
lower panel to a general disc inclined by $70^{\circ}$. Based on the
curves of three reasonable values of $h$, we find that the S/N
required to determine the scale height is extremely dependent on the
disc inclination. For example, if we compare the profile of $h$=0.08
and 0.03, the largest difference in the normalized intensity profiles
is $\sim4\%$ at PA=0$^{\circ}$ and $\sim10\%$ when comparing $h$=0.08
and 0.15.  This can give a rough estimate of the required S/N to
recover the original $h$ in a real observation, i.e. S/N$\gtrsim
1/0.04=25$ to distinguish cases with $h$ between 0.03 and 0.08, and
S/N$\gtrsim 1/0.1=10$ to exclude $h$ of 0.15 or higher. On the other
hand, if the disc is inclined by 70$^{\circ}$ the S/N required to
distinguished between the three cases would be $\gtrsim
1/0.2=5$. Hence, high S/N ALMA observations are fundamental to
constrain the scale height of low inclined debris rings, and thus, the
level of stirring of the debris.

Interestingly, changing $\Delta r$ can produce a similar effect to the
intensity profile. In the bottom panel, the dashed yellow line
corresponding to $h=0.08$ and $\Delta r=10$ AU has a similar profile
compared with the case $h=0.15$ and $\Delta r=23$ AU. However, $\Delta
r$ can be directly estimated measuring the width of the ring along the
major axis, breaking this degeneracy. The anticorrelation between $h$
and $\Delta r$ mentioned in Sec. \ref{sec:model_dust} that appears in
Figure \ref{fig:mcmc} can be qualitatively understood as both $h$ or
$\Delta r$ have an impact on the width of the ring projected in the
sky.

Using the antenna configurations for cycle 3 provided in the CASA
software, we simulate ALMA observations to estimate the necessary
angular resolution to recover an aspect ratio $h$ of 0.08 with an
uncertainty less than 10\% of our best-fit model for HD~181327, using
our RADMC3D-MCMC approach. We find that provided with a maximum
projected baseline of 1.3 km at 220 GHz, i.e. an angular resolution of
$\sim0\farcs25$, and a S/N of $\sim50$ along the ring, it would be
possible to constrain $\langle h\rangle=0.083\pm0.005$.

\section{Summary}
\label{sec:summary}
We resolved the HD~181327 debris ring in dust continuum for the first
time at millimetre wavelengths with an angular resolution of
$0\farcs47\times0\farcs36$ and we detected $^{12}$CO (2-1) emission
for the first time in a debris disc around a solar-type star. Assuming
an axisymmetric disc, we found that the dust continuum is best fitted
with a ring of radius $86\pm0.4$ AU and width $23.1\pm1.0$ AU. At this
angular resolution, the vertical mass distribution is hard to
constrain, but we were able to put an upper limit of 0.14 for the
vertical aspect ratio $h$.

The ALMA observations are consistent with an axisymmetric ring and no
significant residuals along the main ring remain after subtracting the
best fit model to the continuum data. We also derived an upper limit
for any dust overdensity along the primary ring. When we compared to
previous HST observations at optical wavelengths we found that the
derived orientations of the disc on the sky are consistent, but the
ring radius derived from our data is significantly smaller. This
result is consistent with grain size segregation due to radiation
pressure.

Additionally, we detected low-level emission that extends beyond the
primary ring, consistent with either an extended halo of dust or a
secondary ring. This result is robust against different image
synthesis methods and is also recovered in the residuals of our
best-fit ring model. Deeper ALMA or ACA observations could help to
study the origin of this emission.

We found that the CO is co-located with the dust, favouring a
secondary origin scenario. Assuming an axisymmetric model we model the
CO emission in the non-LTE regime with different kinetic gas
temperatures and collisional partner densities. With gas kinetic
temperatures of 50~K and electron densities similar to the ones found
in $\beta$~Pic, the emission is best fitted with a total CO gas mass
of $1.8\pm0.6\times10^{-6}\mathrm{M}_{\oplus}$. In addition, we
derived CO+CO$_2$ ice abundances in planetesimals and we found that
they are consistent with the composition observed in Solar system
comets. Furthermore, it is unlikely that the effect of hydrodynamics
can affect the structure of the disc.

Finally, we showed that ALMA observations can be very useful to
constrain the vertical distribution of discs. The S/N required is
strongly dependent on the disc inclination in the sky, making it
harder for discs close to a face on orientation as HD~181327.


\section*{Acknowledgements}

 We thank Pablo Roman for his help developing the tools uvmem and
 uvsim used in this work. We also thank the referee for a constructive
 report. This paper makes use of the following ALMA data:
 ADS/JAO.ALMA\#2012.1.00437.S and ADS/JAO.ALMA\#2013.1.00523.S. ALMA
 is a partnership of ESO (representing its member states), NSF (USA)
 and NINS (Japan), together with NRC (Canada) and NSC and ASIAA
 (Taiwan) and KASI (Republic of Korea), in cooperation with the
 Republic of Chile. The Joint ALMA Observatory is operated by ESO,
 AUI/NRAO and NAOJ. This work was supported by the European Union
 through ERC grant number 279973. SM, SC, SP acknowledge financial
 support from Millennium Nucleus RC130007 (Chilean Ministry of
 Economy), and additionally by FONDECYT grants 1130949 and
 3140601. GMK is supported by the Royal Society as a Royal Society
 University Research Fellow.




\bibliographystyle{mnras}
\bibliography{SM_pformation} 



\appendix

\section{Posterior distribution of disk parameters}

\begin{figure*}
  \includegraphics[trim=0.0cm 0.0cm 0.0cm 0.0cm, clip=true, width=0.8\textwidth]{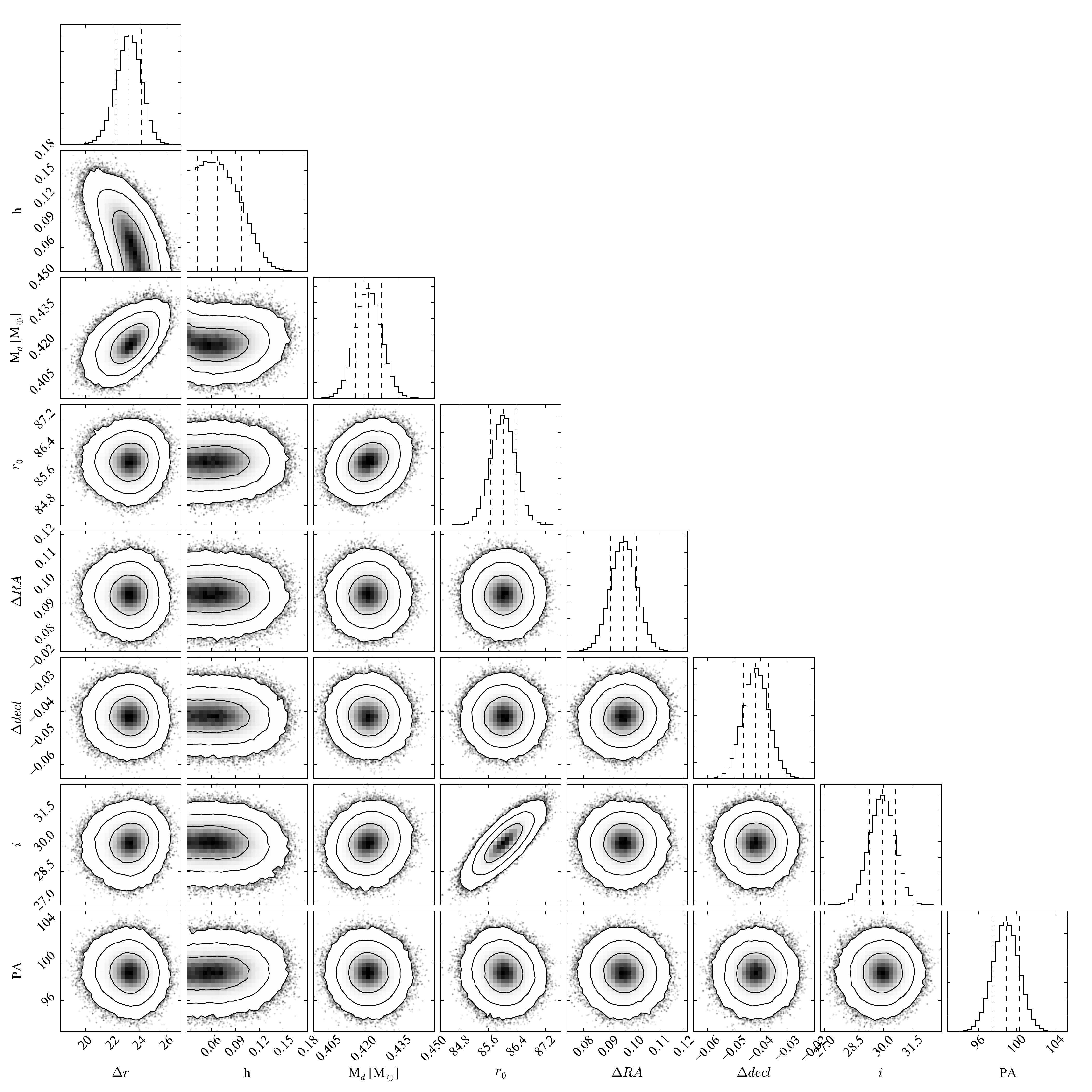}
  \caption{Posterior distribution of $\Delta r$, $h=\Hr/r$, M$_d$,
    $r_0$, RA and decl offsets, $i$ and PA. The 1D marginalised
    distributions are presented in the diagonal, while the 2D
    marginalised distributions are presented in in the bottom left
    half. The vertical dashed lines represent the 16th, 50th and
    84th percentiles. Contours correspond to 68\%, 95\% and 99.7\%
    confidence regions and the black dots to points of the
      MCMC outside the 99.7\% confidence region. This plot was
    generated using the python module \textit{corner}
    \citep{cornerplot}. }
    \label{fig:mcmc_all}
\end{figure*}

In Figure \ref{fig:mcmc_all} we present the posterior distribution of
all the free parameters in the disk model of the dust continuum
described in Sec. \ref{sec:model_dust}.

\bsp	
\label{lastpage}
\end{document}